# Search for Gravitational-wave Signals Associated with Gamma-Ray Bursts during the Second Observing Run of Advanced LIGO and Advanced Virgo

B. P. Abbott[1], R. Abbott[1], T. D. Abbott[2], S. Abraham[3], F. Acernese[4,5], K. Ackley[6], C. Adams[7], R. X. Adhikari[1], V. B. Adya[8], C. Affeldt[9,10], M. Agathos[11,12], K. Agatsuma[13], N. Aggarwal[14], O. D. Aguiar[15], L. Aiello[16,17], A. Ain[3], P. Ajith[18], G. Allen[19], A. Allocca[20,21], M. A. Aloy[22], P. A. Altin[8], A. Amato[23], S. Anand[1], A. Ananyeva[1], S. B. Anderson[1], W. G. Anderson[24], S. V. Angelova[25], S. Antier[26], S. Appert[1], K. Arai[1], M. C. Araya[1], J. S. Areeda[27], M. Arène[26], N. Arnaud[28,29], S. M. Aronson[30], S. Ascenzi[16,31], G. Ashton[6], S. M. Aston[7], P. Astone[32], F. Aubin[33], P. Aufmuth[10], K. AultONeal[34], C. Austin[2], V. Avendano[35], A. Avila-Alvarez[27], S. Babak[26], P. Bacon[26], F. Badaracco[16,17], M. K. M. Bader[36], S. Bae[37], J. Baird[26], P. T. Baker[38], F. Baldaccini[39,40], G. Ballardin[29], S. W. Ballmer[41], A. Bals[34], S. Banagiri[42], J. C. Barayoga[1], C. Barbieri[43,44], S. E. Barclay[45], B. C. Barish[1], D. Barker[46], K. Barkett[47], S. Barnum[14], F. Barone[5,48], B. Barr[45], L. Barsotti[14], M. Barsuglia[26], D. Barta[49], J. Bartlett[46], I. Bartos[30], R. Bassiri[50], A. Basti[20,21], M. Bawaj[40,51], J. C. Bayley[45], M. Bazzan[52,53], B. Bécsy[54], M. Bejger[26,55], I. Belahcene[28], A. S. Bell[45], D. Beniwal[56], M. G. Benjamin[34], B. K. Berger[50], G. Bergmann[9,10], S. Bernuzzi[11], C. P. L. Berry[57], D. Bersanetti[58], A. Bertolini[36], J. Betzwieser[7], R. Bhandare[59], J. Bidler[27], E. Biggs[24], I. A. Bilenko[60], S. A. Bilgili[38], G. Billingsley[1], R. Birney[25], O. Birnholtz[61], S. Biscans[1,14], M. Bischi[62,63], S. Biscoveanu[14], A. Bisht[10], M. Bitossi[21,29], M. A. Bizouard[64], J. K. Blackburn[1], J. Blackman[47], C. D. Blair[7], D. G. Blair[65], R. M. Blair[46], S. Bloemen[66], F. Bobba[67,68], N. Bode[9,10], M. Boer[64], Y. Boetzel[69], G. Bogaert[64], F. Bondu[70], R. Bonnand[33], P. Booker[9,10], B. A. Boom[36], R. Bork[1], V. Boschi[29], S. Bose[3], V. Bossilkov[65], J. Bosveld[65], Y. Bouffanais[52,53], A. Bozzi[29], C. Bradaschia[21], P. R. Brady[24], A. Bramley[7], M. Branchesi[16,17], J. E. Brau[71], M. Breschi[11], T. Briant[72], J. H. Briggs[45], F. Brighenti[62,63], A. Brillet[64], M. Brinkmann[9,10], P. Brockill[24], A. F. Brooks[1], J. Brooks[29], D. D. Brown[56], S. Brunett[1], A. Buikema[14], T. Bulik[73], H. J. Bulten[36,74], A. Buonanno[75,76], D. Buskulic[33], C. Buy[26], R. L. Byer[50], M. Cabero[9,10], L. Cadonati[77], G. Cagnoli[78], C. Cahillane[1], J. Calderón Bustillo[6], T. A. Callister[1], E. Calloni[5,79], J. B. Camp[80], W. A. Campbell[6], M. Canepa[58,81], K. C. Cannon[82], H. Cao[56], J. Cao[83], G. Carapella[67,68], F. Carbognani[29], S. Caride[84], M. F. Carney[57], G. Carullo[20,21], J. Casanueva Diaz[21], C. Casentini[31,85], S. Caudill[36], M. Cavaglià[86,87], F. Cavalier[28], R. Cavalieri[29], G. Cella[21], P. Cerdá-Durán[22], E. Cesarini[31,88], O. Chaibi[64], K. Chakravarti[3], S. J. Chamberlin[89], M. Chan[45], S. Chao[90], P. Charlton[91], E. A. Chase[57], E. Chassande-Mottin[26], D. Chatterjee[24], M. Chaturvedi[59], B. D. Cheeseboro[38], H. Y. Chen[92], X. Chen[65], Y. Chen[47], H.-P. Cheng[30], C. K. Cheong[93], H. Y. Chia[30], F. Chiadini[68,94], A. Chincarini[58], A. Chiummo[29], G. Cho[95], H. S. Cho[96], M. Cho[76], N. Christensen[64,97], Q. Chu[65], S. Chua[72], K. W. Chung[93], S. Chung[65], G. Ciani[52,53], M. Cieślar[55], A. A. Ciobanu[56], R. Ciolfi[53,98], F. Cipriano[64], A. Cirone[58,81], F. Clara[46], J. A. Clark[77], P. Clearwater[99], F. Cleva[64], E. Coccia[16,17], P.-F. Cohadon[72], D. Cohen[28], M. Colleoni[100], C. G. Collette[101], C. Collins[13], M. Colpi[43,44], L. R. Cominsky[102], M. Constancio, Jr.[15], L. Conti[53], S. J. Cooper[13], P. Corban[7], T. R. Corbitt[2], I. Cordero-Carrión[103], S. Corezzi[39,40], K. R. Corley[104], N. Cornish[54], D. Corre[28], A. Corsi[84], S. Cortese[29], C. A. Costa[15], R. Cotesta[75], M. W. Coughlin[1], S. B. Coughlin[57,105], J.-P. Coulon[64], S. T. Countryman[104], P. Couvares[1], P. B. Covas[100], E. E. Cowan[77], D. M. Coward[65], M. J. Cowart[7], D. C. Coyne[1], R. Coyne[106], J. D. E. Creighton[24], T. D. Creighton[107], J. Cripe[2], M. Croquette[72], S. G. Crowder[108], T. J. Cullen[2], A. Cumming[45], L. Cunningham[45], E. Cuoco[29], T. Dal Canton[80], G. Dálya[109], B. D'Angelo[58,81], S. L. Danilishin[9,10], S. D'Antonio[31], K. Danzmann[9,10], A. Dasgupta[110], C. F. Da Silva Costa[30], L. E. H. Datrier[45], V. Dattilo[29], I. Dave[59], M. Davier[28], D. Davis[41], E. J. Daw[111], D. DeBra[50], M. Deenadayalan[3], J. Degallaix[23], M. De Laurentis[5,79], S. Deléglise[72], W. Del Pozzo[20,21], L. M. DeMarchi[57], N. Demos[14], T. Dent[112], R. De Pietri[113,114], R. De Rosa[5,79], C. De Rossi[23,29], R. DeSalvo[115], O. de Varona[9,10], S. Dhurandhar[3], M. C. Díaz[107], T. Dietrich[36], L. Di Fiore[5], C. DiFronzo[13], C. Di Giorgio[67,68], F. Di Giovanni[22], M. Di Giovanni[116,117], T. Di Girolamo[5,79], A. Di Lieto[20,21], B. Ding[101], S. Di Pace[32,118], I. Di Palma[32,118], F. Di Renzo[20,21], A. K. Divakarla[30], A. Dmitriev[13], Z. Doctor[92], F. Donovan[14], K. L. Dooley[86,105], S. Doravari[3], I. Dorrington[105], T. P. Downes[24], M. Drago[16,17], J. C. Driggers[46], Z. Du[83], J.-G. Ducoin[28], P. Dupej[45], O. Durante[67,68], S. E. Dwyer[46], P. J. Easter[6], G. Eddolls[45], T. B. Edo[111], A. Effler[7], P. Ehrens[1], J. Eichholz[8], S. S. Eikenberry[30], M. Eisenmann[33], R. A. Eisenstein[14], L. Errico[5,79], R. C. Essick[92], H. Estelles[100], D. Estevez[33], Z. B. Etienne[38], T. Etzel[1], M. Evans[14], T. M. Evans[7], V. Fafone[16,31,85], S. Fairhurst[105], X. Fan[83], S. Farinon[58], B. Farr[71], W. M. Farr[13], E. J. Fauchon-Jones[105], M. Favata[35], M. Fays[111], M. Fazio[119], C. Fee[120], J. Feicht[1], M. M. Fejer[50], F. Feng[26], A. Fernandez-Galiana[14], I. Ferrante[20,21], E. C. Ferreira[15], T. A. Ferreira[15], F. Fidecaro[20,21], I. Fiori[29], D. Fiorucci[16,17], M. Fishbach[92], R. P. Fisher[121], J. M. Fishner[14], R. Fittipaldi[68,122], M. Fitz-Axen[42], V. Fiumara[68,123], R. Flaminio[33,124], M. Fletcher[45], E. Floden[42], E. Flynn[27], H. Fong[82], J. A. Font[22,125], P. W. F. Forsyth[8], J.-D. Fournier[64], Francisco Hernandez Vivanco[6], S. Frasca[32,118], F. Frasconi[21], Z. Frei[109], A. Freise[13], R. Frey[71], V. Frey[28], P. Fritschel[14], V. V. Frolov[7], G. Fronzè[126], P. Fulda[30], M. Fyffe[7], H. A. Gabbard[45], B. U. Gadre[75], S. M. Gaebel[13], J. R. Gair[127], L. Gammaitoni[39], S. G. Gaonkar[3], C. García-Quirós[100], F. Garufi[5,79], B. Gateley[46], S. Gaudio[34], G. Gaur[128], V. Gayathri[129], G. Gemme[58], E. Genin[29], A. Gennai[21], D. George[19], J. George[59], L. Gergely[130], S. Ghonge[77], Abhirup Ghosh[75], Archisman Ghosh[36], S. Ghosh[24], B. Giacomazzo[116,117], J. A. Giaime[2,7], K. D. Giardina[7], D. R. Gibson[131], K. Gill[104], L. Glover[132], J. Gniesmer[133], P. Godwin[89], E. Goetz[46], R. Goetz[30], B. Goncharov[6], G. González[2], J. M. Gonzalez Castro[20,21], A. Gopakumar[134], S. E. Gossan[1], M. Gosselin[20,21,29], R. Gouaty[33], B. Grace[8], A. Grado[5,135], M. Granata[23], A. Grant[45], S. Gras[14], P. Grassia[1], C. Gray[46], R. Gray[45], G. Greco[62,63], A. C. Green[30], R. Green[105], E. M. Gretarsson[34], A. Grimaldi[116,117], S. J. Grimm[16,17], P. Groot[66], H. Grote[105],

S. Grunewald[75], P. Gruning[28], G. M. Guidi[62,63], H. K. Gulati[110], Y. Guo[36], A. Gupta[89], Anchal Gupta[1], P. Gupta[36], E. K. Gustafson[1], R. Gustafson[136], L. Haegel[100], O. Halim[16,17], B. R. Hall[137], E. D. Hall[14], E. Z. Hamilton[105], G. Hammond[45], M. Haney[69], M. M. Hanke[9,10], J. Hanks[46], C. Hanna[89], M. D. Hannam[105], O. A. Hannuksela[93], T. J. Hansen[34], J. Hanson[7], T. Harder[64], T. Hardwick[2], K. Haris[18], J. Harms[16,17], G. M. Harry[138], I. W. Harry[139], R. K. Hasskew[7], C. J. Haster[14], K. Haughian[45], F. J. Hayes[45], J. Healy[61], A. Heidmann[72], M. C. Heintze[7], H. Heitmann[64], F. Hellman[140], P. Hello[28], G. Hemming[29], M. Hendry[45], I. S. Heng[45], J. Hennig[9,10], M. Heurs[9,10], S. Hild[45], T. Hinderer[36,141,142], S. Hochheim[9,10], D. Hofman[23], A. M. Holgado[19], N. A. Holland[8], K. Holt[7], D. E. Holz[92], P. Hopkins[105], C. Horst[24], J. Hough[45], E. J. Howell[65], C. G. Hoy[105], Y. Huang[14], M. T. Hübner[6], E. A. Huerta[19], D. Huet[28], B. Hughey[34], V. Hui[33], S. Husa[100], S. H. Huttner[45], T. Huynh-Dinh[7], B. Idzkowski[73], A. Iess[31,85], H. Inchauspe[30], C. Ingram[56], R. Inta[84], G. Intini[32,118], B. Irwin[120], H. N. Isa[45], J.-M. Isac[72], M. Isi[14], B. R. Iyer[18], T. Jacqmin[72], S. J. Jadhav[143], K. Jani[77], N. N. Janthalur[143], P. Jaranowski[144], D. Jariwala[30], A. C. Jenkins[145], J. Jiang[30], D. S. Johnson[19], A. W. Jones[13], D. I. Jones[146], J. D. Jones[46], R. Jones[45], R. J. G. Jonker[36], L. Ju[65], J. Junker[9,10], C. V. Kalaghatgi[105], V. Kalogera[57], B. Kamai[1], S. Kandhasamy[3], G. Kang[37], J. B. Kanner[1], S. J. Kapadia[24], S. Karki[71], R. Kashyap[18], M. Kasprzack[1], S. Katsanevas[29], E. Katsavounidis[14], W. Katzman[7], S. Kaufer[10], K. Kawabe[46], N. V. Keerthana[3], F. Kéfélian[64], D. Keitel[139], R. Kennedy[111], J. S. Key[147], F. Y. Khalili[60], I. Khan[16,31], S. Khan[9,10], E. A. Khazanov[148], N. Khetan[16,17], M. Khursheed[59], N. Kijbunchoo[8], Chunglee Kim[149], J. C. Kim[150], K. Kim[93], W. Kim[56], W. S. Kim[151], Y.-M. Kim[152], C. Kimball[57], P. J. King[46], M. Kinley-Hanlon[45], R. Kirchhoff[9,10], J. S. Kissel[46], L. Kleybolte[133], J. H. Klika[24], S. Klimenko[30], T. D. Knowles[38], P. Koch[9,10], S. M. Koehlenbeck[9,10], G. Koekoek[36,153], S. Koley[36], V. Kondrashov[1], A. Kontos[154], N. Koper[9,10], M. Korobko[133], W. Z. Korth[1], M. Kovalam[65], D. B. Kozak[1], C. Krämer[9,10], V. Kringel[9,10], N. Krishnendu[155], A. Królak[156,157], N. Krupinski[24], G. Kuehn[9,10], A. Kumar[143], P. Kumar[158], Rahul Kumar[46], Rakesh Kumar[110], L. Kuo[90], A. Kutynia[156], S. Kwang[24], B. D. Lackey[75], D. Laghi[20,21], K. H. Lai[93], T. L. Lam[93], M. Landry[46], B. B. Lane[14], R. N. Lang[159], J. Lange[61], B. Lantz[50], R. K. Lanza[14], A. Lartaux-Vollard[28], P. D. Lasky[6], M. Laxen[7], A. Lazzarini[1], C. Lazzaro[53], P. Leaci[32,118], S. Leavey[9,10], Y. K. Lecoeuche[46], C. H. Lee[96], H. K. Lee[160], H. M. Lee[161], H. W. Lee[150], J. Lee[95], K. Lee[45], J. Lehmann[9,10], A. K. Lenon[38], N. Leroy[28], N. Letendre[33], Y. Levin[6], A. Li[93], J. Li[83], K. J. L. Li[93], T. G. F. Li[93], X. Li[47], F. Lin[6], F. Linde[36,162], S. D. Linker[132], T. B. Littenberg[163], J. Liu[65], X. Liu[24], M. Llorens-Monteagudo[22], R. K. L. Lo[1,93], L. T. London[14], A. Longo[164,165], M. Lorenzini[16,17], V. Loriette[166], M. Lormand[7], G. Losurdo[21], J. D. Lough[9,10], C. O. Lousto[61], G. Lovelace[27], M. E. Lower[167], H. Lück[9,10], D. Lumaca[31,85], A. P. Lundgren[139], R. Lynch[14], Y. Ma[47], R. Macas[105], S. Macfoy[25], M. MacInnis[14], D. M. Macleod[105], A. Macquet[64], I. Magaña Hernandez[24], F. Magaña-Sandoval[30], R. M. Magee[89], E. Majorana[32], I. Maksimovic[166], A. Malik[59], N. Man[64], V. Mandic[42], V. Mangano[32,45,118], G. L. Mansell[14,46], M. Manske[24], M. Mantovani[29], M. Mapelli[52,53], F. Marchesoni[40,51], F. Marion[33], S. Márka[104], Z. Márka[104], C. Markakis[19], A. S. Markosyan[50], A. Markowitz[1], E. Maros[1], A. Marquina[103], S. Marsat[26], F. Martelli[62,63], I. W. Martin[45], R. M. Martin[35], V. Martinez[78], D. V. Martynov[13], H. Masalehdan[133], K. Mason[14], E. Massera[111], A. Masserot[33], T. J. Massinger[1], M. Masso-Reid[45], S. Mastrogiovanni[26], A. Matas[75], F. Matichard[1,14], L. Matone[104], N. Mavalvala[14], J. J. McCann[65], R. McCarthy[46], D. E. McClelland[8], S. McCormick[7], L. McCuller[14], S. C. McGuire[168], C. McIsaac[139], J. McIver[1], D. J. McManus[8], T. McRae[8], S. T. McWilliams[38], D. Meacher[24], G. D. Meadors[6], M. Mehmet[9,10], A. K. Mehta[18], J. Meidam[36], E. Mejuto Villa[68,115], A. Melatos[99], G. Mendell[46], R. A. Mercer[24], L. Mereni[23], K. Merfeld[71], E. L. Merilh[46], M. Merzougui[64], S. Meshkov[1], C. Messenger[45], C. Messick[89], F. Messina[43,44], R. Metzdorff[72], P. M. Meyers[99], F. Meylahn[9,10], A. Miani[116,117], H. Miao[13], C. Michel[23], H. Middleton[99], L. Milano[5,79], A. L. Miller[30,32,118], M. Millhouse[99], J. C. Mills[105], M. C. Milovich-Goff[132], O. Minazzoli[64,169], Y. Minenkov[31], A. Mishkin[30], C. Mishra[170], T. Mistry[111], S. Mitra[3], V. P. Mitrofanov[60], G. Mitselmakher[30], R. Mittleman[14], G. Mo[97], D. Moffa[120], K. Mogushi[86], S. R. P. Mohapatra[14], M. Molina-Ruiz[140], M. Mondin[132], M. Montani[62,63], C. J. Moore[13], D. Moraru[46], F. Morawski[55], G. Moreno[46], S. Morisaki[82], B. Mours[33], C. M. Mow-Lowry[13], F. Muciaccia[32,118], Arunava Mukherjee[9,10], D. Mukherjee[24], S. Mukherjee[107], Subroto Mukherjee[110], N. Mukund[3,9,10], A. Mullavey[7], J. Munch[56], E. A. Muñiz[41], M. Muratore[34], P. G. Murray[45], I. Nardecchia[31,85], L. Naticchioni[32,118], R. K. Nayak[171], B. F. Neil[65], J. Neilson[68,115], G. Nelemans[36,66], T. J. N. Nelson[7], M. Nery[9,10], A. Neunzert[136], L. Nevin[1], K. Y. Ng[14], S. Ng[56], C. Nguyen[26], P. Nguyen[71], D. Nichols[36,141], S. A. Nichols[2], S. Nissanke[36,141], F. Nocera[29], C. North[105], L. K. Nuttall[139], M. Obergaulinger[22,172], J. Oberling[46], B. D. O'Brien[30], G. Oganesyan[16,17], G. H. Ogin[173], J. J. Oh[151], S. H. Oh[151], F. Ohme[9,10], H. Ohta[82], M. A. Okada[15], M. Oliver[100], P. Oppermann[9,10], Richard J. Oram[7], B. O'Reilly[7], R. G. Ormiston[42], L. F. Ortega[30], R. O'Shaughnessy[61], S. Ossokine[75], D. J. Ottaway[56], H. Overmier[7], B. J. Owen[84], A. E. Pace[89], G. Pagano[20,21], M. A. Page[65], G. Pagliaroli[16,17], A. Pai[129], S. A. Pai[59], J. R. Palamos[71], O. Palashov[148], C. Palomba[32], H. Pan[90], P. K. Panda[143], P. T. H. Pang[36,93], C. Pankow[57], F. Pannarale[32,118], B. C. Pant[59], F. Paoletti[21], A. Paoli[29], A. Parida[3], W. Parker[7,168], D. Pascucci[36,45], A. Pasqualetti[29], R. Passaquieti[20,21], D. Passuello[21], M. Patil[157], B. Patricelli[20,21], E. Payne[6], B. L. Pearlstone[45], T. C. Pechsiri[30], A. J. Pedersen[41], M. Pedraza[1], R. Pedurand[23,174], A. Pele[7], S. Penn[175], A. Perego[116,117], C. J. Perez[46], C. Périgois[33], A. Perreca[116,117], J. Petermann[133], H. P. Pfeiffer[75], M. Phelps[9,10], K. S. Phukon[3], O. J. Piccinni[32,118], M. Pichot[64], F. Piergiovanni[62,63], V. Pierro[68,115], G. Pillant[29], L. Pinard[23], I. M. Pinto[68,88,115], M. Pirello[46], M. Pitkin[45], W. Plastino[164,165], R. Poggiani[20,21], D. Y. T. Pong[93], S. Ponrathnam[3], P. Popolizio[29], J. Powell[167], A. K. Prajapati[110], J. Prasad[3], K. Prasai[50], R. Prasanna[143], G. Pratten[100], T. Prestegard[24], M. Principe[68,88,115], G. A. Prodi[116,117], L. Prokhorov[13], M. Punturo[40], P. Puppo[32], M. Pürrer[75], H. Qi[105], V. Quetschke[107], P. J. Quinonez[34], F. J. Raab[46], G. Raaijmakers[36,141], H. Radkins[46], N. Radulesco[64], P. Raffai[109], S. Raja[59], C. Rajan[59], B. Rajbhandari[84], M. Rakhmanov[107], K. E. Ramirez[107], A. Ramos-Buades[100], Javed Rana[3], K. Rao[57], P. Rapagnani[32,118], V. Raymond[105], M. Razzano[20,21], J. Read[27],

T. Regimbau[33], L. Rei[58], S. Reid[25], D. H. Reitze[1,30], P. Rettegno[126,176], F. Ricci[32,118], C. J. Richardson[34], J. W. Richardson[1], P. M. Ricker[19], G. Riemenschneider[126,176], K. Riles[136], M. Rizzo[57], N. A. Robertson[1,45], F. Robinet[28], A. Rocchi[31], L. Rolland[33], J. G. Rollins[1], V. J. Roma[71], M. Romanelli[70], R. Romano[4,5], C. L. Romel[46], J. H. Romie[7], C. A. Rose[24], D. Rose[27], K. Rose[120], D. Rosińska[73], S. G. Rosofsky[19], M. P. Ross[177], S. Rowan[45], A. Rüdiger[9,10,196], P. Ruggi[29], G. Rutins[131], K. Ryan[46], S. Sachdev[89], T. Sadecki[46], M. Sakellariadou[145], O. S. Salafia[43,44,178], L. Salconi[29], M. Saleem[155], A. Samajdar[36], L. Sammut[6], E. J. Sanchez[1], L. E. Sanchez[1], N. Sanchis-Gual[179], J. R. Sanders[180], K. A. Santiago[35], E. Santos[64], N. Sarin[6], B. Sassolas[23], B. S. Sathyaprakash[89,105], O. Sauter[33,136], R. L. Savage[46], P. Schale[71], M. Scheel[47], J. Scheuer[57], P. Schmidt[13,66], R. Schnabel[133], R. M. S. Schofield[71], A. Schönbeck[133], E. Schreiber[9,10], B. W. Schulte[9,10], B. F. Schutz[105], J. Scott[45], S. M. Scott[8], E. Seidel[19], D. Sellers[7], A. S. Sengupta[181], N. Sennett[75], D. Sentenac[29], V. Sequino[58], A. Sergeev[148], Y. Setyawati[9,10], D. A. Shaddock[8], T. Shaffer[46], M. S. Shahriar[57], M. B. Shaner[132], A. Sharma[16,17], P. Sharma[59], P. Shawhan[76], H. Shen[19], R. Shink[182], D. H. Shoemaker[14], D. M. Shoemaker[77], K. Shukla[140], S. ShyamSundar[59], K. Siellez[77], M. Sieniawska[55], D. Sigg[46], L. P. Singer[80], D. Singh[89], N. Singh[73], A. Singhal[16,32], A. M. Sintes[100], S. Sitmukhambetov[107], V. Skliris[105], B. J. J. Slagmolen[8], T. J. Slaven-Blair[65], J. R. Smith[27], R. J. E. Smith[6], S. Somala[183], E. J. Son[151], S. Soni[2], B. Sorazu[45], F. Sorrentino[58], T. Souradeep[3], E. Sowell[84], A. P. Spencer[45], M. Spera[52,53], A. K. Srivastava[110], V. Srivastava[41], K. Staats[57], C. Stachie[64], M. Standke[9,10], D. A. Steer[26], M. Steinke[9,10], J. Steinlechner[45,133], S. Steinlechner[133], D. Steinmeyer[9,10], S. P. Stevenson[167], D. Stocks[50], R. Stone[107], D. J. Stops[13], K. A. Strain[45], G. Stratta[63,184], S. E. Strigin[60], A. Strunk[46], R. Sturani[185], A. L. Stuver[186], V. Sudhir[14], T. Z. Summerscales[187], L. Sun[1], S. Sunil[110], A. Sur[55], J. Suresh[82], P. J. Sutton[105], B. L. Swinkels[36], M. J. Szczepańczyk[34], M. Tacca[36], S. C. Tait[45], C. Talbot[6], D. B. Tanner[30], D. Tao[1], M. Tápai[130], A. Tapia[27], J. D. Tasson[97], R. Taylor[1], R. Tenorio[100], L. Terkowski[133], M. Thomas[7], P. Thomas[46], S. R. Thondapu[59], K. A. Thorne[7], E. Thrane[6], Shubhanshu Tiwari[116,117], Srishti Tiwari[134], V. Tiwari[105], K. Toland[45], M. Tonelli[20,21], Z. Tornasi[45], A. Torres-Forné[188], C. I. Torrie[1], D. Töyrä[13], F. Travasso[29,40], G. Traylor[7], M. C. Tringali[73], A. Tripathee[136], A. Trovato[26], L. Trozzo[21,189], K. W. Tsang[36], M. Tse[14], R. Tso[47], L. Tsukada[82], D. Tsuna[82], T. Tsutsui[82], D. Tuyenbayev[107], K. Ueno[82], D. Ugolini[190], C. S. Unnikrishnan[134], A. L. Urban[2], S. A. Usman[92], H. Vahlbruch[10], G. Vajente[1], G. Valdes[2], M. Valentini[116,117], N. van Bakel[36], M. van Beuzekom[36], J. F. J. van den Brand[36,74], C. Van Den Broeck[36,191], D. C. Vander-Hyde[41], L. van der Schaaf[36], J. V. VanHeijningen[65], A. A. van Veggel[45], M. Vardaro[52,53], V. Varma[47], S. Vass[1], M. Vasúth[49], A. Vecchio[13], G. Vedovato[53], J. Veitch[45], P. J. Veitch[56], K. Venkateswara[177], G. Venugopalan[1], D. Verkindt[33], F. Vetrano[62,63], A. Viceré[62,63], A. D. Viets[24], S. Vinciguerra[13], D. J. Vine[131], J.-Y. Vinet[64], S. Vitale[14], T. Vo[41], H. Vocca[39,40], C. Vorvick[46], S. P. Vyatchanin[60], A. R. Wade[1], L. E. Wade[120], M. Wade[120], R. Walet[36], M. Walker[27], L. Wallace[1], S. Walsh[24], H. Wang[13], J. Z. Wang[136], S. Wang[19], W. H. Wang[107], Y. F. Wang[93], R. L. Ward[8], Z. A. Warden[34], J. Warner[46], M. Was[33], J. Watchi[101], B. Weaver[46], L.-W. Wei[9,10], M. Weinert[9,10], A. J. Weinstein[1], R. Weiss[14], F. Wellmann[9,10], L. Wen[65], E. K. Wessel[19], P. Weßels[9,10], J. W. Westhouse[34], K. Wette[8], J. T. Whelan[61], B. F. Whiting[30], C. Whittle[14], D. M. Wilken[9,10], D. Williams[45], A. R. Williamson[36,141], J. L. Willis[1], B. Willke[9,10], W. Winkler[9,10], C. C. Wipf[1], H. Wittel[9,10], G. Woan[45], J. Woehler[9,10], J. K. Wofford[61], J. L. Wright[45], D. S. Wu[9,10], D. M. Wysocki[61], S. Xiao[1], R. Xu[108], H. Yamamoto[1], C. C. Yancey[76], L. Yang[119], Y. Yang[30], Z. Yang[42], M. J. Yap[8], M. Yazback[30], D. W. Yeeles[105], Hang Yu[14], Haocun Yu[14], S. H. R. Yuen[93], A. K. Zadrożny[107], A. Zadrożny[156], M. Zanolin[34], T. Zelenova[29], J.-P. Zendri[53], M. Zevin[57], J. Zhang[65], L. Zhang[1], T. Zhang[45], C. Zhao[65], G. Zhao[101], M. Zhou[57], Z. Zhou[57], X. J. Zhu[6], M. E. Zucker[1,14], J. Zweizig[1]

The LIGO Scientific Collaboration and the Virgo Collaboration,

and

R. L. Aptekar[192], W. V. Boynton[193], D. D. Frederiks[192], S. V. Golenetskii[192], D. V. Golovin[194], K. Hurley[195], A. V. Kozlova[192], M. L. Litvak[194], I. G. Mitrofanov[194], A. B. Sanin[194], D. S. Svinkin[192]

IPN Collaboration,

Francesco Carotenuto[32,118], and Badri Krishnan[9,10]

[1] LIGO, California Institute of Technology, Pasadena, CA 91125, USA; lsc-spokesperson@ligo.org, virgo-spokesperson@ego-gw.it
[2] Louisiana State University, Baton Rouge, LA 70803, USA
[3] Inter-University Centre for Astronomy and Astrophysics, Pune 411007, India
[4] Dipartimento di Farmacia, Università di Salerno, I-84084 Fisciano, Salerno, Italy
[5] INFN, Sezione di Napoli, Complesso Universitario di Monte S.Angelo, I-80126 Napoli, Italy
[6] OzGrav, School of Physics & Astronomy, Monash University, Clayton 3800, Victoria, Australia
[7] LIGO Livingston Observatory, Livingston, LA 70754, USA
[8] OzGrav, Australian National University, Canberra, Australian Capital Territory 0200, Australia
[9] Max Planck Institute for Gravitational Physics (Albert Einstein Institute), D-30167 Hannover, Germany
[10] Leibniz Universität Hannover, D-30167 Hannover, Germany
[11] Theoretisch-Physikalisches Institut, Friedrich-Schiller-Universität Jena, D-07743 Jena, Germany
[12] University of Cambridge, Cambridge CB2 1TN, UK
[13] University of Birmingham, Birmingham B15 2TT, UK
[14] LIGO, Massachusetts Institute of Technology, Cambridge, MA 02139, USA
[15] Instituto Nacional de Pesquisas Espaciais, 12227-010 São José dos Campos, São Paulo, Brazil
[16] Gran Sasso Science Institute (GSSI), I-67100 L'Aquila, Italy
[17] INFN, Laboratori Nazionali del Gran Sasso, I-67100 Assergi, Italy
[18] International Centre for Theoretical Sciences, Tata Institute of Fundamental Research, Bengaluru 560089, India
[19] NCSA, University of Illinois at Urbana-Champaign, Urbana, IL 61801, USA
[20] Università di Pisa, I-56127 Pisa, Italy
[21] INFN, Sezione di Pisa, I-56127 Pisa, Italy

[22] Departamento de Astronomía y Astrofísica, Universitat de València, E-46100 Burjassot, València, Spain
[23] Laboratoire des Matériaux Avancés (LMA), CNRS/IN2P3, F-69622 Villeurbanne, France
[24] University of Wisconsin-Milwaukee, Milwaukee, WI 53201, USA
[25] SUPA, University of Strathclyde, Glasgow G1 1XQ, UK
[26] APC, AstroParticule et Cosmologie, Université Paris Diderot, CNRS/IN2P3, CEA/Irfu, Observatoire de Paris, Sorbonne Paris Cité, F-75205 Paris Cedex 13, France
[27] California State University Fullerton, Fullerton, CA 92831, USA
[28] LAL, Univ. Paris-Sud, CNRS/IN2P3, Université Paris-Saclay, F-91898 Orsay, France
[29] European Gravitational Observatory (EGO), I-56021 Cascina, Pisa, Italy
[30] University of Florida, Gainesville, FL 32611, USA
[31] INFN, Sezione di Roma Tor Vergata, I-00133 Roma, Italy
[32] INFN, Sezione di Roma, I-00185 Roma, Italy
[33] Laboratoire d'Annecy de Physique des Particules (LAPP), Univ. Grenoble Alpes, Université Savoie Mont Blanc, CNRS/IN2P3, F-74941 Annecy, France
[34] Embry-Riddle Aeronautical University, Prescott, AZ 86301, USA
[35] Montclair State University, Montclair, NJ 07043, USA
[36] Nikhef, Science Park 105, 1098 XG Amsterdam, The Netherlands
[37] Korea Institute of Science and Technology Information, Daejeon 34141, Republic of Korea
[38] West Virginia University, Morgantown, WV 26506, USA
[39] Università di Perugia, I-06123 Perugia, Italy
[40] INFN, Sezione di Perugia, I-06123 Perugia, Italy
[41] Syracuse University, Syracuse, NY 13244, USA
[42] University of Minnesota, Minneapolis, MN 55455, USA
[43] Università degli Studi di Milano-Bicocca, I-20126 Milano, Italy
[44] INFN, Sezione di Milano-Bicocca, I-20126 Milano, Italy
[45] SUPA, University of Glasgow, Glasgow G12 8QQ, UK
[46] LIGO Hanford Observatory, Richland, WA 99352, USA
[47] Caltech CaRT, Pasadena, CA 91125, USA
[48] Dipartimento di Medicina, Chirurgia e Odontoiatria "Scuola Medica Salernitana," Università di Salerno, I-84081 Baronissi, Salerno, Italy
[49] Wigner RCP, RMKI, H-1121 Budapest, Konkoly Thege Miklós út 29-33, Hungary
[50] Stanford University, Stanford, CA 94305, USA
[51] Università di Camerino, Dipartimento di Fisica, I-62032 Camerino, Italy
[52] Università di Padova, Dipartimento di Fisica e Astronomia, I-35131 Padova, Italy
[53] INFN, Sezione di Padova, I-35131 Padova, Italy
[54] Montana State University, Bozeman, MT 59717, USA
[55] Nicolaus Copernicus Astronomical Center, Polish Academy of Sciences, 00-716, Warsaw, Poland
[56] OzGrav, University of Adelaide, Adelaide, South Australia 5005, Australia
[57] Center for Interdisciplinary Exploration & Research in Astrophysics (CIERA), Northwestern University, Evanston, IL 60208, USA
[58] INFN, Sezione di Genova, I-16146 Genova, Italy
[59] RRCAT, Indore, Madhya Pradesh 452013, India
[60] Faculty of Physics, Lomonosov Moscow State University, Moscow 119991, Russia
[61] Rochester Institute of Technology, Rochester, NY 14623, USA
[62] Università degli Studi di Urbino "Carlo Bo," I-61029 Urbino, Italy
[63] INFN, Sezione di Firenze, I-50019 Sesto Fiorentino, Firenze, Italy
[64] Artemis, Université Côte d'Azur, Observatoire Côte d'Azur, CNRS, CS 34229, F-06304 Nice Cedex 4, France
[65] OzGrav, University of Western Australia, Crawley, Western Australia 6009, Australia
[66] Department of Astrophysics/IMAPP, Radboud University Nijmegen, P.O. Box 9010, 6500 GL Nijmegen, The Netherlands
[67] Dipartimento di Fisica "E.R. Caianiello," Università di Salerno, I-84084 Fisciano, Salerno, Italy
[68] INFN, Sezione di Napoli, Gruppo Collegato di Salerno, Complesso Universitario di Monte S. Angelo, I-80126 Napoli, Italy
[69] Physik-Institut, University of Zurich, Winterthurerstrasse 190, 8057 Zurich, Switzerland
[70] Univ Rennes, CNRS, Institut FOTON - UMR6082, F-3500 Rennes, France
[71] University of Oregon, Eugene, OR 97403, USA
[72] Laboratoire Kastler Brossel, Sorbonne Université, CNRS, ENS-Université PSL, Collège de France, F-75005 Paris, France
[73] Astronomical Observatory Warsaw University, 00-478 Warsaw, Poland
[74] VU University Amsterdam, 1081 HV Amsterdam, The Netherlands
[75] Max Planck Institute for Gravitational Physics (Albert Einstein Institute), D-14476 Potsdam-Golm, Germany
[76] University of Maryland, College Park, MD 20742, USA
[77] School of Physics, Georgia Institute of Technology, Atlanta, GA 30332, USA
[78] Université de Lyon, Université Claude Bernard Lyon 1, CNRS, Institut Lumière Matière, F-69622 Villeurbanne, France
[79] Università di Napoli "Federico II," Complesso Universitario di Monte S.Angelo, I-80126 Napoli, Italy
[80] NASA Goddard Space Flight Center, Greenbelt, MD 20771, USA
[81] Dipartimento di Fisica, Università degli Studi di Genova, I-16146 Genova, Italy
[82] RESCEU, University of Tokyo, Tokyo, 113-0033, Japan
[83] Tsinghua University, Beijing 100084, People's Republic of China
[84] Texas Tech University, Lubbock, TX 79409, USA
[85] Università di Roma Tor Vergata, I-00133 Roma, Italy
[86] The University of Mississippi, University, MS 38677, USA
[87] Missouri University of Science and Technology, Rolla, MO 65409, USA
[88] Museo Storico della Fisica e Centro Studi e Ricerche "Enrico Fermi," I-00184 Roma, Italy
[89] The Pennsylvania State University, University Park, PA 16802, USA
[90] National Tsing Hua University, Hsinchu City, 30013 Taiwan, Republic of China
[91] Charles Sturt University, Wagga Wagga, New South Wales 2678, Australia
[92] University of Chicago, Chicago, IL 60637, USA
[93] The Chinese University of Hong Kong, Shatin, NT, Hong Kong
[94] Dipartimento di Ingegneria Industriale (DIIN), Università di Salerno, I-84084 Fisciano, Salerno, Italy
[95] Seoul National University, Seoul 08826, Republic of Korea
[96] Pusan National University, Busan 46241, Republic of Korea

[97] Carleton College, Northfield, MN 55057, USA
[98] INAF, Osservatorio Astronomico di Padova, I-35122 Padova, Italy
[99] OzGrav, University of Melbourne, Parkville, Victoria 3010, Australia
[100] Universitat de les Illes Balears, IAC3—IEEC, E-07122 Palma de Mallorca, Spain
[101] Université Libre de Bruxelles, Brussels B-1050, Belgium
[102] Sonoma State University, Rohnert Park, CA 94928, USA
[103] Departamento de Matemáticas, Universitat de València, E-46100 Burjassot, València, Spain
[104] Columbia University, New York, NY 10027, USA
[105] Cardiff University, Cardiff CF24 3AA, UK
[106] University of Rhode Island, Kingston, RI 02881, USA
[107] The University of Texas Rio Grande Valley, Brownsville, TX 78520, USA
[108] Bellevue College, Bellevue, WA 98007, USA
[109] MTA-ELTE Astrophysics Research Group, Institute of Physics, Eötvös University, Budapest 1117, Hungary
[110] Institute for Plasma Research, Bhat, Gandhinagar 382428, India
[111] The University of Sheffield, Sheffield S10 2TN, UK
[112] IGFAE, Campus Sur, Universidade de Santiago de Compostela, E-15782, Spain
[113] Dipartimento di Scienze Matematiche, Fisiche e Informatiche, Università di Parma, I-43124 Parma, Italy
[114] INFN, Sezione di Milano Bicocca, Gruppo Collegato di Parma, I-43124 Parma, Italy
[115] Dipartimento di Ingegneria, Università del Sannio, I-82100 Benevento, Italy
[116] Università di Trento, Dipartimento di Fisica, I-38123 Povo, Trento, Italy
[117] INFN, Trento Institute for Fundamental Physics and Applications, I-38123 Povo, Trento, Italy
[118] Università di Roma "La Sapienza," I-00185 Roma, Italy
[119] Colorado State University, Fort Collins, CO 80523, USA
[120] Kenyon College, Gambier, OH 43022, USA
[121] Christopher Newport University, Newport News, VA 23606, USA
[122] CNR-SPIN, c/o Università di Salerno, I-84084 Fisciano, Salerno, Italy
[123] Scuola di Ingegneria, Università della Basilicata, I-85100 Potenza, Italy
[124] National Astronomical Observatory of Japan, 2-21-1 Osawa, Mitaka, Tokyo 181-8588, Japan
[125] Observatori Astronòmic, Universitat de València, E-46980 Paterna, València, Spain
[126] INFN Sezione di Torino, I-10125 Torino, Italy
[127] School of Mathematics, University of Edinburgh, Edinburgh EH9 3FD, UK
[128] Institute of Advanced Research, Gandhinagar 382426, India
[129] Indian Institute of Technology Bombay, Powai, Mumbai 400 076, India
[130] University of Szeged, Dóm tér 9, Szeged 6720, Hungary
[131] SUPA, University of the West of Scotland, Paisley PA1 2BE, UK
[132] California State University, Los Angeles, 5151 State University Dr., Los Angeles, CA 90032, USA
[133] Universität Hamburg, D-22761 Hamburg, Germany
[134] Tata Institute of Fundamental Research, Mumbai 400005, India
[135] INAF, Osservatorio Astronomico di Capodimonte, I-80131 Napoli, Italy
[136] University of Michigan, Ann Arbor, MI 48109, USA
[137] Washington State University, Pullman, WA 99164, USA
[138] American University, Washington, DC 20016, USA
[139] University of Portsmouth, Portsmouth, PO1 3FX, UK
[140] University of California, Berkeley, CA 94720, USA
[141] GRAPPA, Anton Pannekoek Institute for Astronomy and Institute for High-Energy Physics, University of Amsterdam, Science Park 904, 1098 XH Amsterdam, The Netherlands
[142] Delta Institute for Theoretical Physics, Science Park 904, 1090 GL Amsterdam, The Netherlands
[143] Directorate of Construction, Services & Estate Management, Mumbai 400094, India
[144] University of Białystok, 15-424 Białystok, Poland
[145] King's College London, University of London, London WC2R 2LS, UK
[146] University of Southampton, Southampton SO17 1BJ, UK
[147] University of Washington Bothell, Bothell, WA 98011, USA
[148] Institute of Applied Physics, Nizhny Novgorod, 603950, Russia
[149] Ewha Womans University, Seoul 03760, Republic of Korea
[150] Inje University Gimhae, South Gyeongsang 50834, Republic of Korea
[151] National Institute for Mathematical Sciences, Daejeon 34047, Republic of Korea
[152] Ulsan National Institute of Science and Technology, Ulsan 44919, Republic of Korea
[153] Maastricht University, P.O. Box 616, 6200 MD Maastricht, The Netherlands
[154] Bard College, 30 Campus Rd., Annandale-On-Hudson, NY 12504, USA
[155] Chennai Mathematical Institute, Chennai 603103, India
[156] NCBJ, 05-400 Świerk-Otwock, Poland
[157] Institute of Mathematics, Polish Academy of Sciences, 00656 Warsaw, Poland
[158] Cornell University, Ithaca, NY 14850, USA
[159] Hillsdale College, Hillsdale, MI 49242, USA
[160] Hanyang University, Seoul 04763, Republic of Korea
[161] Korea Astronomy and Space Science Institute, Daejeon 34055, Republic of Korea
[162] Institute for High-Energy Physics, University of Amsterdam, Science Park 904, 1098 XH Amsterdam, The Netherlands
[163] NASA Marshall Space Flight Center, Huntsville, AL 35811, USA
[164] Dipartimento di Matematica e Fisica, Università degli Studi Roma Tre, I-00146 Roma, Italy
[165] INFN, Sezione di Roma Tre, I-00146 Roma, Italy
[166] ESPCI, CNRS, F-75005 Paris, France
[167] OzGrav, Swinburne University of Technology, Hawthorn VIC 3122, Australia
[168] Southern University and A&M College, Baton Rouge, LA 70813, USA
[169] Centre Scientifique de Monaco, 8 quai Antoine Ier, MC-98000, Monaco
[170] Indian Institute of Technology Madras, Chennai 600036, India
[171] IISER-Kolkata, Mohanpur, West Bengal 741252, India

[172] Institut für Kernphysik, Theoriezentrum, D-64289 Darmstadt, Germany
[173] Whitman College, 345 Boyer Avenue, Walla Walla, WA 99362, USA
[174] Université de Lyon, F-69361 Lyon, France
[175] Hobart and William Smith Colleges, Geneva, NY 14456, USA
[176] Dipartimento di Fisica, Università degli Studi di Torino, I-10125 Torino, Italy
[177] University of Washington, Seattle, WA 98195, USA
[178] INAF, Osservatorio Astronomico di Brera sede di Merate, I-23807 Merate, Lecco, Italy
[179] Centro de Astrofísica e Gravitação (CENTRA), Departamento de Física, Instituto Superior Técnico, Universidade de Lisboa, 1049-001 Lisboa, Portugal
[180] Marquette University, 11420 W. Clybourn St., Milwaukee, WI 53233, USA
[181] Indian Institute of Technology, Gandhinagar Ahmedabad Gujarat 382424, India
[182] Université de Montréal/Polytechnique, Montreal, QC H3T 1J4, Canada
[183] Indian Institute of Technology Hyderabad, Sangareddy, Khandi, Telangana 502285, India
[184] INAF, Osservatorio di Astrofisica e Scienza dello Spazio, I-40129 Bologna, Italy
[185] International Institute of Physics, Universidade Federal do Rio Grande do Norte, Natal RN 59078-970, Brazil
[186] Villanova University, 800 Lancaster Ave., Villanova, PA 19085, USA
[187] Andrews University, Berrien Springs, MI 49104, USA
[188] Max Planck Institute for Gravitationalphysik (Albert Einstein Institute), D-14476 Potsdam-Golm, Germany
[189] Università di Siena, I-53100 Siena, Italy
[190] Trinity University, San Antonio, TX 78212, USA
[191] Van Swinderen Institute for Particle Physics and Gravity, University of Groningen, Nijenborgh 4, 9747 AG Groningen, The Netherlands
[192] Ioffe Institute, Politekhnicheskaya 26, St. Petersburg 194021, Russia
[193] Lunar and Planetary Laboratory, University of Arizona, Tucson, AZ, USA
[194] Space Research Institute, Russian Academy of Sciences, Moscow 117997, Russia
[195] University of California, Berkeley, Space Sciences Laboratory, 7 Gauss Way, Berkeley, CA 94720-7450, USA

## Abstract

We present the results of targeted searches for gravitational-wave transients associated with gamma-ray bursts during the second observing run of Advanced LIGO and Advanced Virgo, which took place from 2016 November to 2017 August. We have analyzed 98 gamma-ray bursts using an unmodeled search method that searches for generic transient gravitational waves and 42 with a modeled search method that targets compact-binary mergers as progenitors of short gamma-ray bursts. Both methods clearly detect the previously reported binary merger signal GW170817, with $p$-values of $<9.38 \times 10^{-6}$ (modeled) and $3.1 \times 10^{-4}$ (unmodeled). We do not find any significant evidence for gravitational-wave signals associated with the other gamma-ray bursts analyzed, and therefore we report lower bounds on the distance to each of these, assuming various source types and signal morphologies. Using our final modeled search results, short gamma-ray burst observations, and assuming binary neutron star progenitors, we place bounds on the rate of short gamma-ray bursts as a function of redshift for $z \leqslant 1$. We estimate 0.07–1.80 joint detections with *Fermi*-GBM per year for the 2019–20 LIGO-Virgo observing run and 0.15–3.90 per year when current gravitational-wave detectors are operating at their design sensitivities.

## 1. Introduction

Gamma-ray bursts (GRBs) are high-energy astrophysical transients originating throughout the universe that are observed more than once per day on average. The prompt gamma-ray emission is thought to emanate from highly relativistic jets powered by matter interacting with a compact central object such as an accreting black hole (BH) or a magnetar (Woosley 1993). Broadly speaking, GRBs are divided into two subpopulations based on duration and spectral hardness (Kouveliotou et al. 1993).

Long-soft bursts generally have durations $\gtrsim 2$ s. The favored model is the core-collapse supernova (SN) of a rapidly rotating massive star (Woosley & Bloom 2006; Mösta et al. 2015). This connection was observationally supported by the presence of SN 1998bw within the error box of the long GRB 980425 (Galama et al. 1998) and the later strong association of SN 2003dh with GRB 030329 (Hjorth et al. 2003; Stanek et al. 2003). The core-collapse process will produce some gravitational radiation (Fryer & New 2011). Rotational instabilities may give rise to much more significant gravitational-wave (GW) emission, however, and could be observable from beyond the Milky Way (Davies et al. 2002; Fryer et al. 2002; Kobayashi & Meszaros 2003; Shibata et al. 2003; Piro & Pfahl 2007; Corsi & Meszaros 2009; Romero et al. 2010; Gossan et al. 2016).

Neutron star (NS) binaries have long been proposed as the progenitors of short-hard GRBs (Blinnikov et al. 1984; Paczynski 1986; Eichler et al. 1989; Narayan et al. 1992). The detection of the GW transient GW170817, an NS binary merger (Abbott et al. 2017a, 2017e, 2019b), in coincidence with the short GRB 170817A (Goldstein et al. 2017; Savchenko et al. 2017), confirmed that such mergers can produce short GRBs. An optical detection of a counterpart (Coulter et al. 2017) was followed by panchromatic observations identifying kilonova and afterglow emission (see Abbott et al. 2017f, and references therein).

The unusually low flux of GRB 170817A and its light-curve evolution suggested an off-axis GRB with a relativistic structured jet or cocoon that either propagated into the universe successfully or was choked (Rossi et al. 2002; Hallinan et al. 2017; Kasliwal et al. 2017; Lamb & Kobayashi 2017; Troja et al. 2017; Gottlieb et al. 2018; Lazzati et al. 2018; Zhang et al. 2018). Later, very long baseline interferometry observations indicated a successfully launched relativistic jet (Mooley et al. 2018; Ghirlanda et al. 2019). The center of this jet appears to have been directed at an angle of approximately 15°–30° from the line of sight (Lazzati et al. 2018; Mooley et al. 2018). Analysis of the first 10 yr of *Fermi* Gamma-ray Burst Monitor (GBM) data suggests that GRB 170817A may belong to a population of local, low-luminosity short GRBs with similar spectral features (von Kienlin et al. 2019). The multimessenger observations of this event have proven to be extremely rich, providing insights about the structure of NSs (Margalit & Metzger 2017; Abbott et al. 2018; De et al. 2018; Most et al. 2018; Radice et al. 2018), the local cosmological expansion rate (Abbott et al. 2017b, 2019b; Hotokezaka et al. 2019), and heavy-element nucleosynthesis (Abbott et al. 2017d; Chornock et al. 2017; Cowperthwaite et al. 2017; Drout et al. 2017; Kasen et al. 2017; Smartt et al. 2017), to name a few.

[196] Deceased, 2018 July.

In this paper we present targeted GW follow-up of GRBs—long and short—reported during the second observing run of Advanced LIGO and Advanced Virgo (O2). The observing run spanned 2016 November 30 to 2017 August 25, with Advanced Virgo commencing observations on 2017 August 1. As a measure of their sensitivities, the Advanced LIGO instruments had sky- and orientation-averaged binary neutron star (BNS) ranges between 65 and 100 Mpc throughout the run, while for Advanced Virgo this range was approximately 25 Mpc (Abbott et al. 2019a). In addition to GW170817, seven binary BH mergers were previously identified during O2, with a further three binary BHs observed during the first observing run (Abbott et al. 2019a).

We discuss the population of GRBs included in our analyses in Section 2 and summarize the methods used in Section 3. We then present the results of a modeled binary merger analysis targeting short-hard GRBs in Section 4 and an unmodeled analysis targeting all GRBs in Section 5, followed by discussion in Section 6 and concluding remarks in Section 7.

## 2. GRB Sample

The GRB sample contains events disseminated by the Gamma-ray Coordinates Network (GCN),[197] with additional information gathered from the *Swift* BAT catalog[198] (Lien et al. 2016), the online *Swift* GRB Archive,[199] the *Fermi* GBM Burst Catalog[200] (Gruber et al. 2014; von Kienlin et al. 2014; Bhat et al. 2016), and the Interplanetary Network (IPN; Hurley et al. 2003).[201] An automated system called VALID (Coyne 2015) cross-checks the time and localization parameters of the *Swift* and *Fermi* events against the published catalog with automated literature searches. In total, from 2016 November through 2017 August, there were 242 bursts detected in the combined *Swift* + *Fermi* catalog. We received a total of 52 bursts localized by the IPN, with many bursts appearing in both catalogs. GRBs that were poorly localized were removed from our sample, as were GRBs that did not occur during a period of stable, science-quality data taken by the available GW detectors.

For the purposes of this work, GRBs are classified (as in Abbott et al. 2017g) based on their $T_{90}$ value—the period over which 90% of the flux was observed—and its uncertainty $\delta T_{90}$. GRBs with a value of $T_{90} + \delta T_{90} < 2$ s are *short*, and those with $T_{90} + \delta T_{90} > 4$ s are *long*. The remaining GRBs are *ambiguous*.

As in Abbott et al. (2017g), a generic unmodeled GW transient search (Sutton et al. 2010; Was et al. 2012) was performed for all GRBs for which 660 s of coincident data was available from two GW detectors, regardless of classification. A modeled search for coalescing binary GW signals (Harry & Fairhurst 2011; Williamson et al. 2014) was performed for all short and ambiguous GRBs with at least 1664 s of data in one or more detectors. This scheme resulted in 98 GRBs being analyzed with our unmodeled method and 42 analyzed with our modeled method.

## 3. Search Methods

To cover all possible GW emission mechanisms, we consider two search methods: a modeled search for binary merger signals from short or ambiguous GRBs, and an unmodeled search for GWs from all GRBs. Neither of these methods has changed since previous published results (Abbott et al. 2017a, 2017g), so we provide summary overviews here.

### 3.1. Modeled Search for Binary Mergers

The modeled search is a coherent matched filtering pipeline known as `PyGRB` (Harry & Fairhurst 2011; Williamson et al. 2014) and is contained within the `PyCBC` data analysis toolkit[202] (Nitz et al. 2018). We analyze a 6 s *on-source* window comprising $[-5, +1)$ s around the arrival time of the GRB for a GW candidate event and up to approximately 90 minutes of adjacent data to characterize the background.

We use a bank of GW template waveforms for filtering (Owen & Sathyaprakash 1999) that encompasses combinations of masses and spins consistent with BNS and NS–BH systems that may be electromagnetically bright, i.e., under conservative assumptions about the NS equation of state, the evolution of these systems toward merger could feasibly produce an accretion disk via disruption of the NS that might be sufficient to power a GRB (Pannarale & Ohme 2014). The templates are restricted to orbital inclinations of 0° or 180°. This decision is motivated by the expectation that short GRBs do not have jets with angular sizes, and therefore inclinations, much greater than 30° (e.g., Fong et al. 2015). The effect of a small inclination angle on the relative amplitudes of the two GW polarizations is minor enough that restricting the inclination of templates to 0° or 180° can simultaneously reduce computational cost and improve sensitivity to slightly inclined systems by lowering the search background (Williamson et al. 2014). The templates are generated with an aligned-spin model tuned to numerical simulations of binary BHs (Khan et al. 2016). This model was chosen since it was found to provide good levels of signal recovery with relatively low computational cost, and all available models featuring matter effects or generic spin orientations would significantly increase the average computational cost per individual waveform generation and require a substantial increase in the number of templates. Filtering is performed over frequencies of 30–1000 Hz.

[197] GCN Circulars Archive: http://gcn.gsfc.nasa.gov/gcn3_archive.html.
[198] *Swift* BAT Gamma-Ray Burst Catalog: http://swift.gsfc.nasa.gov/results/batgrbcat/.
[199] *Swift* GRB Archive: http://swift.gsfc.nasa.gov/archive/grb_table/.
[200] FERMIGBRST—*Fermi* GBM Burst Catalog: https://heasarc.gsfc.nasa.gov/W3Browse/fermi/fermigbrst.html.
[201] Collected via private communication with Kevin Hurley.
[202] https://pycbc.org/

The detection statistic is a reweighted, coherent matched filter signal-to-noise ratio (S/N; Harry & Fairhurst 2011; Williamson et al. 2014). Candidate significance is evaluated by comparing the most prominent trigger within the 6 s on-source, if there is one, with the most prominent in each of the numerous 6 s off-source trials to produce a *p*-value for the on-source candidate. Extended background characterization is achieved using *time slides*; additional off-source trials are generated by combining data from GW detectors after introducing time shifts longer than the light-travel time across the network.

Search sensitivity is estimated by injecting simulated signals into off-source data in software. We choose three distinct astrophysical populations of simulated signals: BNS, NS–BH with spins aligned with the orbital angular momentum, and NS–BH with generically oriented spins. Signals are simulated as having originated at a range of distances. The 90% exclusion distance, $D_{90}$, is the distance within which 90% of a simulated population is recovered with a ranking statistic greater than the most significant trigger in the on-source.

In all instances NS masses are drawn from a normal distribution of mean 1.4 $M_{\odot}$ and standard deviation 0.2 $M_{\odot}$ (Kiziltan et al. 2013; Özel & Freire 2016), restricted to the range [1, 3] $M_{\odot}$, where the upper limit is conservatively chosen based on theoretical consideration (Kalogera & Baym 1996). NS spin magnitudes are limited to $\leqslant 0.4$ based on the fastest observed pulsar spin (Hessels et al. 2006).

BH masses are drawn from a normal distribution of mean 10 $M_{\odot}$ and standard deviation 6 $M_{\odot}$, restricted to the range [3, 15] $M_{\odot}$, with spin magnitudes restricted to $\leqslant 0.98$, motivated by X-ray binary observations (e.g., Özel et al. 2010; Kreidberg et al. 2012; Miller & Miller 2014).

All simulations have binary orbital inclinations $\theta_{JN}$, defined as the angle between the total angular momentum and the line of sight, drawn uniformly in $\sin\theta_{JN}$, where $\theta_{JN}$ is restricted to the ranges [0°, 30°] and [150°, 180°].

Additionally, the EM-bright condition is applied to simulations, avoiding the inclusion of systems that could not feasibly power a GRB (Pannarale & Ohme 2014).

For each of our three astrophysical populations we generate simulations with three different waveform models so as to account for modeling uncertainty. Specifically, the results quoted in this paper are obtained for simulations with a point-particle effective one body model tuned to numerical simulations, which incorporates orbital precession effects due to unaligned spins (Pan et al. 2014; Taracchini et al. 2014; Babak et al. 2017).

### *3.2. Unmodeled Search for Generic Transients*

We run an unmodeled search targeting all GRBs; long, short, and ambiguous. This analysis is implemented within the `X-Pipeline` software package (Sutton et al. 2010; Was et al. 2012). This is an unmodeled search since we do not know the specific signal shape of GW emission from the core collapse of massive stars, so we make minimal assumptions about the signal morphology. We use the time interval around a GRB trigger beginning 600 s before and ending either 60 s after or at the $T_{90}$ time (whichever is larger) as the on-source window. This window is long enough to cover the time delay between GW emission from a progenitor and the GRB (Koshut et al. 1995; Aloy et al. 2000; MacFadyen et al. 2001; Zhang et al. 2003; Lazzati 2005; Wang & Meszaros 2007; Burlon et al. 2008, 2009; Lazzati et al. 2009; Vedrenne & Atteia 2009). We restrict the search to the most sensitive frequency band of the GW detectors of 20–500 Hz. At lower frequencies terrestrial noise dominates, and at higher frequencies ($f \gtrsim 300$) the GW energy necessary to produce a detectable signal scales as $\propto f^4$ Hz (see, e.g., Section 2 of Abbott et al. 2017c).

Before analyzing detector data, we excise periods of poor-quality data from the data stream. These periods include non-Gaussian noise transients, or *glitches*, that can be traced to environmental or instrumental causes (Berger 2018; Nuttall 2018). Including a detector data stream with low sensitivity and many glitches can reduce overall search sensitivity. Particular care was taken to ensure that periods of poor-quality data from the Virgo detector, which was significantly less sensitive than both LIGO detectors during O2, did not degrade the unmodeled search performance. For GRBs for which we have data from three interferometers, methods for flagging and removing poor-quality data were tuned on off-source Virgo data; however, ultimately Virgo data were only included in the final analysis if the sensitivity of the search was improved by their inclusion.

The analysis pipeline generates time–frequency maps of the GW data stream after coherently combining data from all detectors. These maps are scanned for clusters of pixels with excess energy, referred to as *events*, which are ranked according to a detection statistic based on energy. Coherent consistency tests are applied to reject events associated with noise transients based on correlations between data in different detectors. The surviving event with the largest ranking statistic is taken to be the best candidate for a GW detection, and we evaluate its significance in the same way as the modeled analysis except with 660 s long off-source trials.

As in the modeled search, we estimate the sensitivity of the unmodeled search by injecting simulated signals into off-source data in software. Here we report results using signals from a stellar collapse model represented by circular sine-Gaussian (CSG) waveforms (see Equation (1) and Section 3.2 of Abbott et al. 2017g), with an optimistic total radiated energy $E_{\rm GW} = 10^{-2}\, M_{\odot}\, c^2$ and fixed $Q$ factor of 9. We construct four sets of such waveforms with central frequencies of 70, 100, 150, and 300 Hz. For an optimistic example of longer-duration GW emission detectable by the unmodeled search, we also report results for five accretion disk instability (ADI) waveforms (van Putten 2001; van Putten et al. 2014). In ADI models, GWs are emitted when instabilities form in a magnetically suspended torus around a rapidly spinning BH. The model specifics and parameters used to generate these ADI models are the same as in both Table 1 and Section 3.2 of Abbott et al. (2017g).

## 4. Modeled Search Results

We analyzed 42 short and ambiguous GRBs with the modeled search during O2. As previously reported, the analysis identifies GW170817 in association with GRB 170817A (Abbott et al. 2017e) in a manner consistent with other GW analyses (Abbott et al. 2017a, 2019b). In our analysis of GRB 170817A reported here, where improved data calibration and noise subtraction have been incorporated, this signal was seen with a measured *p*-value of $<9.38 \times 10^{-6}$ and a coherent S/N of 31.26, far in excess of the loudest background.

**Table 1**
Median 90% Confidence Level Exclusion Distances, $D_{90}$, for the Searches during O2

| Modeled Search (Short GRBs) | BNS | NS–BH Generic Spins | NS–BH Aligned Spins |
|---|---|---|---|
| $D_{90}$ (Mpc) | 80 | 105 | 144 |

| Unmodeled Search (All GRBs) | CSG 70 Hz | CSG 100 Hz | CSG 150 Hz | CSG 300 Hz |
|---|---|---|---|---|
| $D_{90}$ (Mpc) | 112 | 113 | 81 | 38 |

| Unmodeled Search (All GRBs) | ADI A | ADI B | ADI C | ADI D | ADI E |
|---|---|---|---|---|---|
| $D_{90}$ (Mpc) | 32 | 104 | 40 | 15 | 36 |

**Note.** Modeled search results are shown for three classes of NS binary progenitor model, and unmodeled search results are shown for CSG (Abbott et al. 2017g) and ADI (van Putten 2001; van Putten et al. 2014) models.

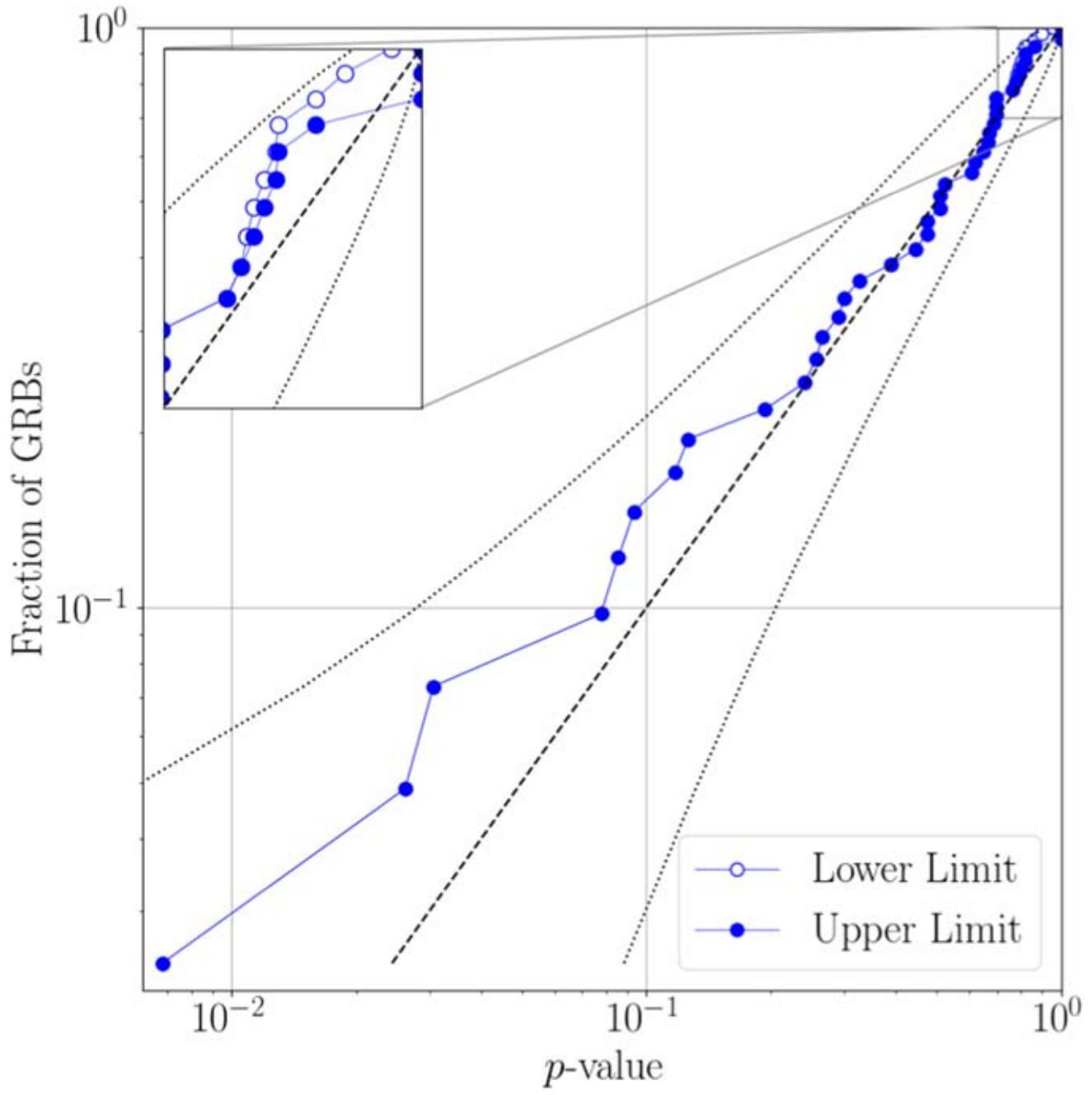

**Figure 1.** Cumulative distribution of event $p$-values for the NS binary search in O2. If the search reports no trigger in the on-source, we plot an upper limit on the $p$-value of 1 and a lower limit equal to the number of off-source trials that contained no trigger. The dashed line indicates the expected distribution of $p$-values under the no-signal hypothesis, with the corresponding $2\sigma$ envelope marked by dotted lines.

We detected no GW signals with significant $p$-values in association with any of the other GRBs. The $p$-value distribution for the 41 GRBs other than GRB 170817A is shown in Figure 1. For GRBs without any associated on-source trigger we plot an upper limit on the $p$-value of 1 and a lower limit given by counting the background trials that similarly had no trigger. The expected distribution under the no-signal hypothesis is shown by the dashed black line, with dotted lines denoting a $2\sigma$ deviation about the no-signal distribution. To quantify population consistency with the no-signal hypothesis, we use the weighted binomial test outlined in Abadie et al. (2012b). This test considers the lowest 5% of $p$-values in the population, weighted by the prior probability of detection based on the detector network sensitivity at the time and in the direction of the GRB. We do not include GW170817, as it is a definite GW detection. This results in a $p$-value of 0.30; thus, we did not find significant evidence for a population of unidentified subthreshold signals with this test.

In addition to GRB 170817A, there were six instances of on-source candidates with $p$-values less than 0.1. The second most significant $p$-value was 0.0068, associated with GRB 170125102 from the *Fermi* GBM burst catalog. These six candidates were the subjects of further data quality checks to assess whether they could be caused by known instrumental noise sources. After careful scrutiny of the data, there were no clear noise artifacts identified as being responsible for any of these candidates. We also ran Bayesian parameter estimation analyses using `LALInference` (Veitch et al. 2015) to quantify the evidence for the presence of a coherent subthreshold NS binary merger signal in the data versus incoherent or Gaussian instrumental noise (Isi et al. 2018). The results of these studies are summarized in more detail in Table 2. In particular, we quote Bayes factors (BFs) to quantify the support for a coherent signal over incoherent or Gaussian noise, where a value less than 1 favors noise over signal and values greater than ~3 are generally required before considering support to be substantial (Kass & Raftery 1995). Some studies have previously looked at the distributions of these BFs in the presence of weak signals and instrumental noise (Veitch & Vecchio 2008; Isi et al. 2018), although in somewhat different contexts to the low-mass targeted coherent search reported here. An in-depth study tailored to this analysis is beyond the scope of this work. However, given that these candidates were initially identified by our coherent matched filter analysis with low S/N, we might expect the BFs to indicate the presence of some degree of coherent power. Our follow-up results reflect this expectation and appear consistent with the search results, with neither significant evidence in favor of incoherent or purely Gaussian noise nor significant evidence in favor of the presence of signals in addition to GW170817 (i.e., $\frac{1}{3} \lesssim \mathrm{BF} \lesssim 3$ in all cases). The largest BF was 2.08 in the case of 170726249 ($p$-value = 0.0262). We also note that, in the absence of a signal with moderate S/N, inferred posterior probability distributions will be prior dominated, and in the presence of non-Gaussian noise fluctuations parameter estimation methods may return broad posteriors with multiple peaks, even for typically well-constrained parameters such as the chirp mass (Huang et al. 2018). We observe these posterior features in our follow-up analyses as noted in Table 2.

GRB 170817A is known to have originated at a distance of ~43 Mpc in the galaxy NGC 4993 (Abbott et al. 2017e). We have plotted the cumulative 90% exclusion distances for the remaining short and ambiguous GRBs in Figure 2. For each of our three simulated signal classes we quote the median of the 41$D_{90}$ results in Table 1.

## 5. Unmodeled Search Results

A total of 98 GRBs were analyzed using the generic transient method, and no significant events were found except for GRB 170817A. The generic method recovered a signal for GRB 170817A consistent with the previously reported signal GW170817 at a $p$-value of $3.1 \times 10^{-4}$. This value differs slightly from that reported in Abbott et al. (2017e), which can be explained by various changes in the configuration of `X-Pipeline`. First, the clustering of pixels in time–frequency maps was previously done over a $7 \times 7$ pixel grid, whereas in

**Table 2**
Results of Follow-up Studies of PyGRB Candidates with $p < 0.1$

| GRB Name | $p$-value | BF | $\hat{\rho}$ | Comment |
|---|---|---|---|---|
| 161210524 | 0.0933 | 1.45 | 6.51 | Weak Bayesian evidence in favor of a coherent signal over noise. Chirp mass posterior is broad with multiple peaks. |
| 170125102 | 0.0068 | 0.88 | 6.23 | Weak Bayesian evidence in favor of noise over a coherent signal. Posteriors show no significant information gain over priors. Chirp mass posterior is broad and multimodal. |
| 170206453 | 0.0418 | 0.94 | 6.89 | Weak Bayesian evidence in favor of noise over a coherent signal. Chirp mass posterior is broad with multiple peaks. |
| 170219002 | 0.0307 | 0.88 | 5.96 | Weak Bayesian evidence in favor of noise over a coherent signal. Posteriors show minimal information gain over priors. Chirp mass posterior is broad with multiple peaks. |
| 170614505 | 0.0856 | 0.46 | 6.43 | Weak Bayesian evidence in favor of noise over a coherent signal. Posteriors show no significant information gain over priors. Chirp mass posterior is broad with multiple peaks. |
| 170726249 | 0.0262 | 2.08 | 6.91 | Weak Bayesian evidence in favor of a coherent signal over noise. Chirp mass posterior is broad with a single peak. |

**Note.** Bayes factors (BFs) quantify the Bayesian odds ratio between the hypothesis that there is a coherent NS binary merger signal in the data and the hypothesis that the data contain only instrumental noise, which may be purely Gaussian or include incoherent non-Gaussianities (see Equation (1) and accompanying discussion in Isi et al. 2018). At low S/N, inferred posterior probability distributions tend to be prior dominated and, in the presence of non-Gaussian noise fluctuations, may exhibit multiple peaks, even for typically well-constrained parameters such as the chirp mass (Huang et al. 2018). We report here $\hat{\rho}$, the network matched filter S/N corresponding to the maximum of the likelihood as estimated by LALInference.

the analysis reported here all clustering is done in a 3 × 3 grid. Second, in the case of GRB 170817A the coherent veto tests were tuned (as described in Section III of Sutton et al. 2010) to maximize the sensitivity of the search to injections of BNS waveforms on the 99.99999th percentile loudest data segment. Here, we go back to the coherent veto tuning used in previous searches that uses the background data segment containing the 95th percentile loudest background event to all injected waveform families.

For the population of results we have compared the distribution of $p$-values against the expected distribution under the no-signal hypothesis, shown in Figure 3. We find a combined $p$-value of 0.75 (0.75 in O1) looking at the most significant 5% of events from the unmodeled search using the weighted binomial test from Abadie et al. (2012a).

For GRBs other than GRB 170817A we place 90% confidence level lower limits on the distance $D_{90}$ assuming various emission models. The distribution of these lower limits for two models, ADI model A (van Putten 2001; van Putten et al. 2014) and a circular sine-Gaussian with central frequency of 150 Hz (Abbott et al. 2017g), is shown in Figure 4. These limits depend on detector sensitivity, which changes over time and sky location; systematic errors due to mismatch of a true GW signal and the waveforms used in simulations; and amplitude and phase errors from detector calibration. In Table 1 we provide population median exclusion limits for each model used, which vary from 15 to 113 Mpc. Some of these limits differ by an order of magnitude owing to our limited knowledge of burst-type source emission models. The median $D_{90}$ values compare favorably with those from the first observing run, either increasing or staying the same depending on the specific signal model.

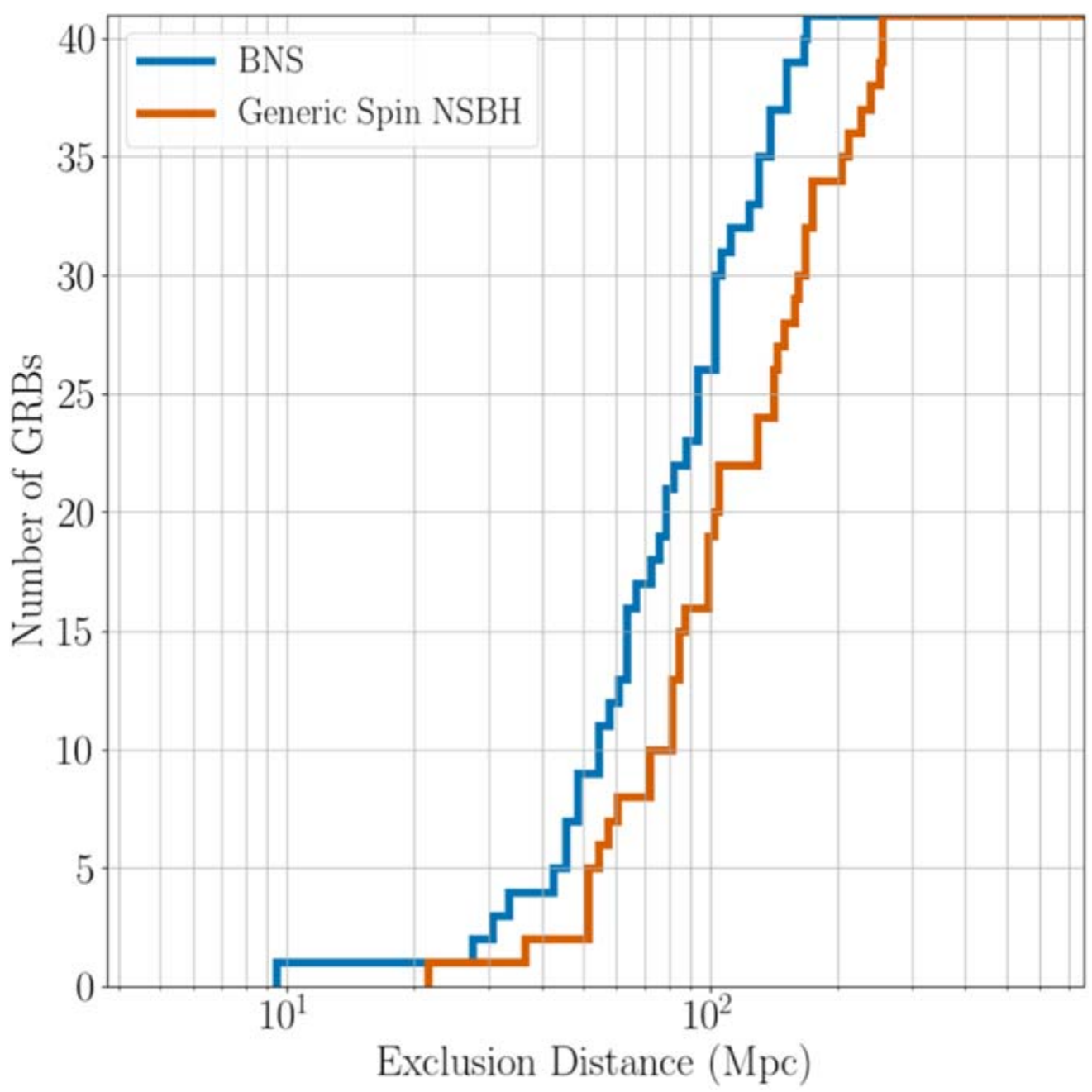

**Figure 2.** Cumulative histograms of the 90% confidence exclusion distances, $D_{90}$, for the BNS (blue) and generically spinning NS–BH (orange) signal models, shown for the sample of 41 short and ambiguous GRBs that did not have an identified GW counterpart. For a given GRB and signal model, $D_{90}$ is the distance within which 90% of simulated signals inserted into off-source data are recovered with greater significance than the most significant on-source trigger. These simulated signals have orbital inclinations $\theta_{JN}$—the angle between the total angular momentum and the line of sight—drawn uniformly in $\sin\theta_{JN}$ with $\theta_{JN}$ restricted to within the ranges [0°, 30°] and [150°, 180°].

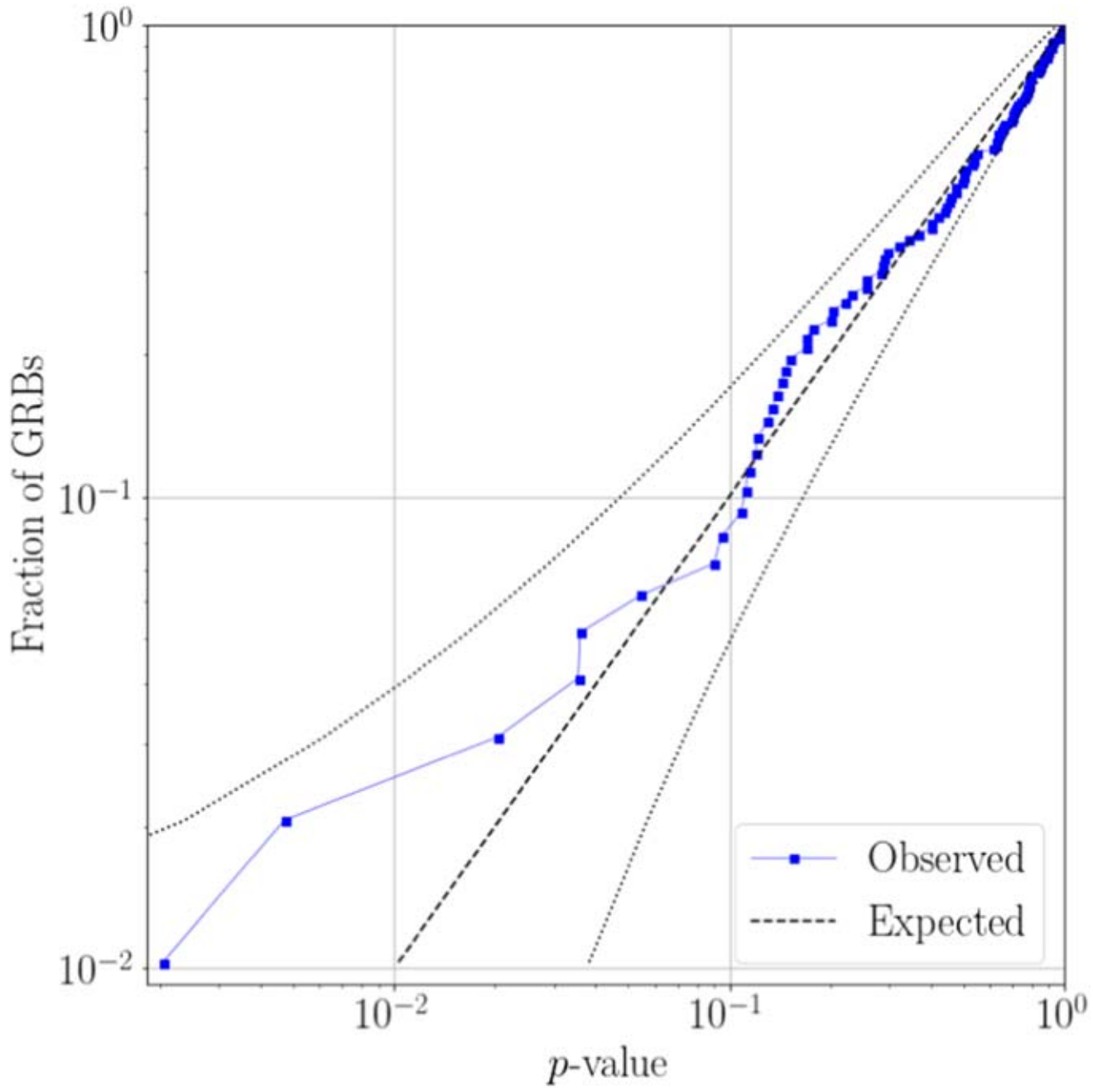

**Figure 3.** Cumulative distribution of $p$-values from the unmodeled search for transient GWs associated with 97 GRBs. The dashed line represents the expected distribution under the no-signal hypothesis, with dotted lines indicating a $2\sigma$ deviation from this distribution. These results are consistent with the no-signal hypothesis and have a combined $p$-value of 0.75 as calculated by a weighted binomial test (Abadie et al. 2012a).

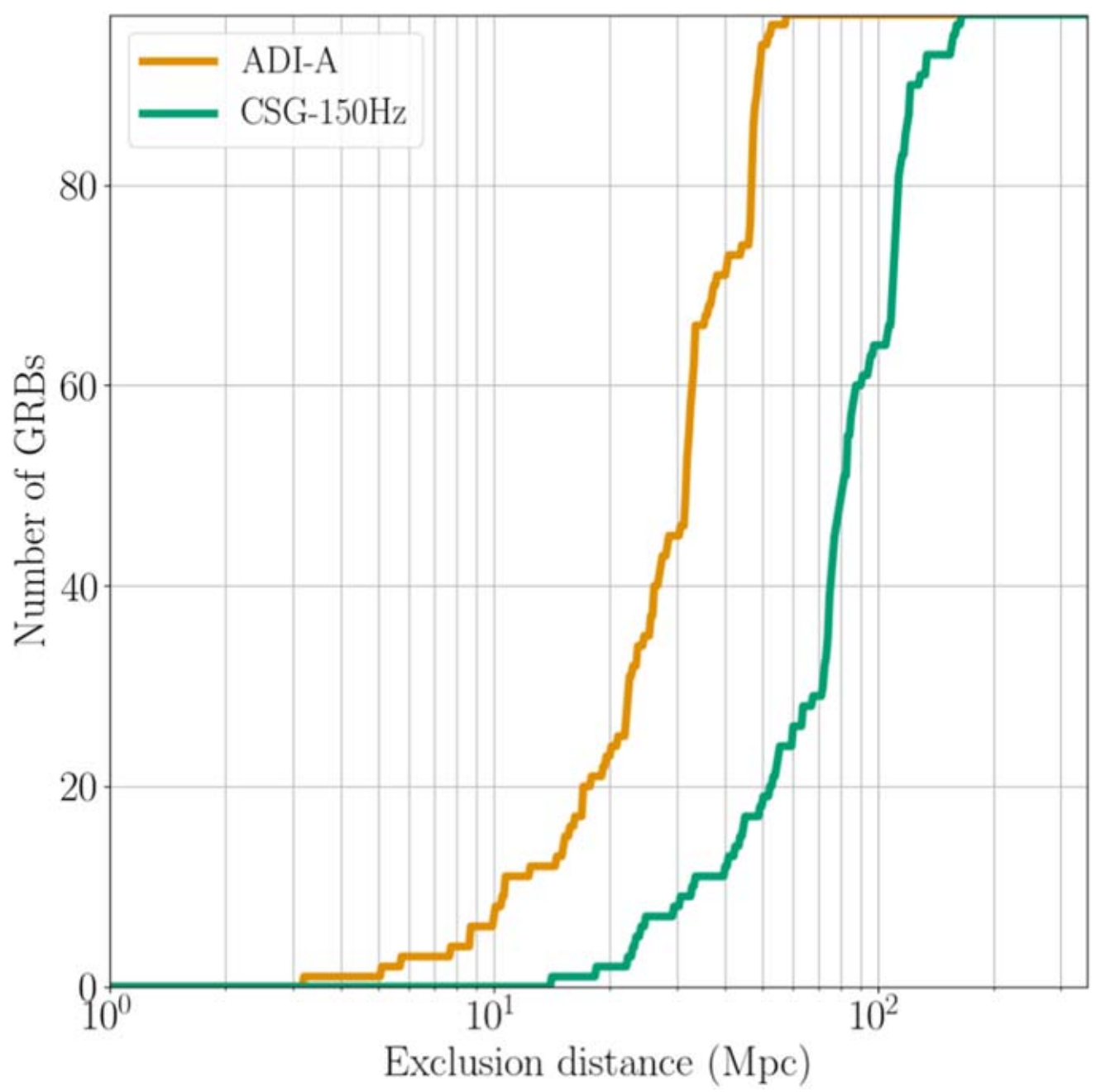

**Figure 4.** Cumulative histograms of the 90% confidence exclusion distances $D_{90}$ for accretion disk instability signal model A (van Putten 2001; van Putten et al. 2014) and the circular sine-Gaussian 150 Hz (Abbott et al. 2017g) model. For a given GRB and signal model this is the distance within which 90% of simulated signals inserted into off-source data are successfully recovered with a significance greater than the loudest on-source trigger. The median values for ADI-A and CSG-150 Hz waveforms are 32 and 81 Mpc, respectively.

## 6. Discussion

Aside from GW170817, no GWs associated with GRBs were detected in O2. The median $D_{90}$ values for each class of signal/source type provide an estimate of roughly how sensitive the searches were to such signals over the course of the entirety of O2, and these are given in Table 1. In Table 3 we provide information on each GRB that was analyzed, including selected $D_{90}$ results where relevant.

The nondetection of GW counterparts for 41 short and ambiguous GRBs analyzed by `PyGRB` can be combined with observed GRBs and the observation of GW170817 to obtain bounds on the short GRB-BNS rate as a function of redshift.

To evaluate this rate given the uncertainty in the jet structure profile of the short-GRB population, we model the GRB luminosity function as a broken power law following Wanderman & Piran (2015), but extended at low luminosities with a second break with an associated free parameter $\gamma_L$, as in Abbott et al. (2017e). This extension at low luminosity is an effective model of the short-GRB jet structure that yields low luminosities for mergers seen at a wide angle from their rotation axis:

$$\phi_o(L_{\rm i}) = \begin{cases} \left(\frac{L_{\rm i}}{L_{\star\star}}\right)^{-\gamma_L}\left(\frac{L_{\star\star}}{L_\star}\right)^{-\alpha_L} & L_{\rm i} < L_{\star\star} \\ \left(\frac{L_{\rm i}}{L_\star}\right)^{-\alpha_L} & L_{\star\star} < L_{\rm i} < L_\star \\ \left(\frac{L_{\rm i}}{L_\star}\right)^{-\beta_L} & L_{\rm i} > L_\star, \end{cases} \quad (1)$$

where $L_{\rm i}$ is the isotropic equivalent energy and the parameters $L_\star \simeq 2 \times 10^{52}\,{\rm erg\,s^{-1}}$, $L_{\star\star} \simeq 5 \times 10^{49}\,{\rm erg\,s^{-1}}$, $\alpha_L \simeq 1$, and $\beta_L \simeq 2$ were used to fit the observed short-GRB redshift distribution. We assume a threshold value for detectability in *Fermi*-GBM of 2 photons $\rm cm^{-2}\,s^{-1}$ for the 64 ms peak photon flux in the 50–300 keV band. Furthermore, we model the short-GRB spectrum using a Band function (Band et al. 1993) with $E_{\rm peak} = 800\,{\rm keV}$, $\alpha_{\rm Band} = -0.5$, and $\beta_{\rm Band} = -2.25$. This yields an observed redshift distribution normalized by a total *Fermi*-GBM detection rate of 40 short GRBs per year.

In order to constrain the free parameter $\gamma_L$, we start with an uninformative prior on $\gamma_L$, which yields a flat prior on the logarithm of the local rate density. Using the redshift distribution for a given $\gamma_L$, we use Monte Carlo sampling to compute the probability of obtaining the O2 results presented here (41 nondetections and a single detection). This yields a posterior on $\gamma_L$ with 90% confidence bounds of [0.04, 0.98]. The corresponding rates as a function of redshift are shown in Figure 5 in magenta.

These bounds can be compared to other measurements and models of the short-GRB redshift distribution. For instance, the sample of observed short-GRB redshifts without GRB 170817A is shown in Figure 5 by the brown lines (Abbott et al. 2017e, and references therein). We also show the cumulative *Fermi* detection rate as a function of redshift in green, calculated following the framework in Howell et al. (2019). This assumes that all short GRBs are associated with BNS mergers and estimates the *Fermi*-GBM detection rate by scaling the BNS source rate evolution with redshift by the *Fermi*-GBM detection efficiency. Finally, the current estimate of the local BNS merger rate of $1210^{+3230}_{-1040}$ $\rm Gpc^{-3}\,yr^{-1}$ (Abbott et al. 2019a) is shown in black for reference. We find that the posterior bounds from the modeled O2 GRB analysis overlap with the BNS merger rate and *Fermi*-GBM-detected short-GRB rate at low redshift. At high redshift there is agreement with the observed short-GRB redshift distribution and the *Fermi*-GBM detection rate.

For the 2019–2020 LIGO-Virgo observing run we expect to see 1–30 BNS coalescences, while at design sensitivity LIGO-Virgo could detect 4–97 BNS mergers per year. Using the framework provided in Howell et al. (2019), we estimate joint GW-GRB detection rates with *Fermi*-GBM of 0.07–1.80 per year for the 2019–2020 LIGO-Virgo observing run and 0.15–3.90 per year at design sensitivity. We note that although the BNS detection rate for LIGO-Virgo at design sensitivity is around three times higher than that of the 2019–2020 observing run, the joint GW-GRB detection

**Table 3**
Information and Limits on Associated GW Emission for Each of the Analyzed GRBs

| | | | | | | | $D_{90}$ (Mpc) | | | | |
|---|---|---|---|---|---|---|---|---|---|---|---|
| GRB Name | UTC Time | R.A. | Decl. | Satellite(s) | Type | Network | BNS | Generic NS–BH | Aligned NS–BH | ADI-A | CSG-150 Hz |
| 161207224 | 05:22:47 | $19^{h}39^{m}14^{s}$ | −9°56′ | *Fermi* | Long | H1L1 | ⋯ | ⋯ | ⋯ | 8 | 40 |
| 161207813 | 19:31:22 | $3^{h}55^{m}09^{s}$ | 15°44′ | *Fermi* | Long | H1L1 | ⋯ | ⋯ | ⋯ | 26 | 73 |
| 161210524 | 12:33:54 | $18^{h}52^{m}28^{s}$ | 63°03′ | *Fermi* | Ambiguous | H1L1 | 61 | 72 | 112 | 19 | 49 |
| 161212652 | 15:38:59 | $01^{h}\,39^{m}36^{s}$ | 68°12′ | *Fermi* | Ambiguous | H1 | 49 | 59 | 60 | ⋯ | ⋯ |
| 161217128 | 03:03:45 | $14^{h}26^{m}31^{s}$ | 51°59′ | *Fermi* | Ambiguous | H1L1 | 65 | 85 | 122 | 18 | 56 |
| 170111815 | 19:34:01 | $18^{h}\,03^{m}31^{s}$ | 63°42′ | *Fermi* | Ambiguous | H1 | 95 | 160 | 198 | ⋯ | ⋯ |
| 170111A | 00:33:27 | $1^{h}22^{m}45^{s}$ | −32°33′ | *Swift* | Long | H1L1 | ⋯ | ⋯ | ⋯ | 13 | 78 |
| 170112A | 02:01:59 | $1^{h}00^{m}55^{s}$ | −17°14′ | *Swift* | Short | H1L1 | 83 | 106 | 144 | 32 | 79 |
| 170113A[a] | 10:04:04 | $4^{h}06^{m}59^{s}$ | −71°56′ | *Swift* | Long | H1L1 | ⋯ | ⋯ | ⋯ | 32 | 107 |
| 170121067 | 01:36:53 | $0^{h}12^{m}07^{s}$ | −75°37′ | *Fermi* | Ambiguous | H1L1 | 79 | 105 | 144 | 26 | 73 |
| 170121133 | 03:10:52 | $16^{h}07^{m}57^{s}$ | 13°49′ | *Fermi* | Ambiguous | H1L1 | 96 | 142 | 172 | 23 | 88 |
| 170124238 | 05:42:12 | $19^{h}26^{m}57^{s}$ | 69°37′ | *Fermi* | Long | H1L1 | ⋯ | ⋯ | ⋯ | 25 | 72 |
| 170124528 | 12:40:29 | $00^{h}\,43^{m}24^{s}$ | 11°01′ | *Fermi* | Short | H1 | 65 | 101 | 116 | ⋯ | ⋯ |
| 170125022 | 00:31:14 | $17^{h}\,36^{m}34^{s}$ | 28°34′ | *Fermi* | Ambiguous | H1 | 46 | 52 | 57 | ⋯ | ⋯ |
| 170125102 | 02:27:10 | $23^{h}57^{m}38^{s}$ | −38°14′ | *Fermi* | Short | H1L1(H1)[b] | 30 | 39 | 63 | 20 | 51 |
| 170127067 | 01:35:47 | $22^{h}37^{m}19^{s}$ | −63°56′ | *Fermi* | Short | H1L1 | 76 | 129 | 141 | 24 | 64 |
| 170127B | 15:13:29 | $01^{h}\,19^{m}58^{s}$ | −30°20′ | *Swift* | Short | H1 | 113 | 169 | 197 | ⋯ | ⋯ |
| 170130302 | 07:14:44 | $18^{h}04^{m}12^{s}$ | −29°07′ | *Fermi* | Long | H1L1 | ⋯ | ⋯ | ⋯ | 48 | 121 |
| 170130510 | 12:13:48 | $20^{h}35^{m}00^{s}$ | 1°26′ | *Fermi* | Long | H1L1$^{\dagger}$ | ⋯ | ⋯ | ⋯ | 26 | 68 |
| 170202A[c] | 18:28:02 | $10^{h}10^{m}06^{s}$ | 5°01′ | *Swift* | Long | H1L1 | ⋯ | ⋯ | ⋯ | 47 | 113 |
| 170203486 | 11:40:25 | $16^{h}20^{m}21^{s}$ | −0°31′ | *Fermi* | Short | H1L1 | 66 | 99 | 119 | 10 | 81 |
| 170203A | 00:03:41 | $22^{h}11^{m}26^{s}$ | 25°11′ | *Swift* | Long | H1L1 | ⋯ | ⋯ | ⋯ | 38 | 112 |
| 170206A | 10:51:58 | $14^{h}\,12^{m}43^{s}$ | 12°34′ | IPN | Short | H1L1 | 151 | 254 | 264 | 50 | 122 |
| 170208553 | 13:16:33 | $18^{h}57^{m}40^{s}$ | −0°07′ | *Fermi* | Long | H1L1 | ⋯ | ⋯ | ⋯ | 31 | 64 |
| 170208A | 18:11:16 | $11^{h}06^{m}10^{s}$ | −46°47′ | *Swift* | Long | H1L1 | ⋯ | ⋯ | ⋯ | 50 | 134 |
| 170208B | 22:33:38 | $8^{h}28^{m}34^{s}$ | −9°02′ | *Swift* | Long | H1L1$^{\dagger}$ | ⋯ | ⋯ | ⋯ | 32 | 77 |
| 170210116 | 02:47:36 | $15^{h}04^{m}14^{s}$ | −65°06′ | *Fermi* | Long | H1L1$^{\dagger}$ | ⋯ | ⋯ | ⋯ | 49 | 122 |
| 170212034 | 00:49:00 | $10^{h}20^{m}24^{s}$ | −1°29′ | *Fermi* | Long | H1L1 | ⋯ | ⋯ | ⋯ | 29 | 76 |
| 170219002 | 00:03:07 | $3^{h}39^{m}21^{s}$ | 50°04′ | *Fermi* | Short | H1L1 | 171 | 251 | 304 | 52 | 159 |
| 170219110 | 02:38:04 | $5^{h}14^{m}45^{s}$ | −41°14′ | *Fermi* | Long | H1L1 | ⋯ | ⋯ | ⋯ | 10 | 33 |
| 170222A | 05:00:59 | $19^{h}\,31^{m}53^{s}$ | 28°04′ | IPN | Short | H1L1 | 80 | 86 | 112 | 23 | 60 |
| 170302166 | 03:58:24 | $10^{h}17^{m}00^{s}$ | 29°23′ | *Fermi* | Ambiguous | H1L1 | 107 | 175 | 206 | 47 | 109 |
| 170304003 | 00:04:26 | $22^{h}02^{m}00^{s}$ | −73°46′ | *Fermi* | Short | H1L1 | 105 | 143 | 178 | 34 | 85 |
| 170305256 | 06:09:06 | $2^{h}34^{m}38^{s}$ | 12°05′ | *Fermi* | Short | H1L1(L1)[d] | 48 | 73 | 82 | 10 | 14 |
| 170306130 | 03:07:17 | $10^{h}31^{m}31^{s}$ | 27°45′ | *Fermi* | Long | H1L1 | ⋯ | ⋯ | ⋯ | 45 | 111 |
| 170310417 | 09:59:50 | $14^{h}33^{m}14^{s}$ | 53°59′ | *Fermi* | Long | H1L1 | ⋯ | ⋯ | ⋯ | 50 | 135 |
| 170310883 | 21:11:43 | $10^{h}26^{m}43^{s}$ | 41°34′ | *Fermi* | Long | H1L1 | ⋯ | ⋯ | ⋯ | 5 | 23 |
| 170311 | 13:45:09 | $23^{h}43^{m}48^{s}$ | 33°24′ | IPN | Long | H1L1 | ⋯ | ⋯ | ⋯ | 34 | 92 |
| 170311A | 08:08:42 | $18^{h}42^{m}09^{s}$ | −30°02′ | *Swift* | Long | H1L1 | ⋯ | ⋯ | ⋯ | 22 | 43 |
| 170317A | 09:45:59 | $6^{h}12^{m}20^{s}$ | 50°30′ | *Swift* | Long | H1L1 | ⋯ | ⋯ | ⋯ | 33 | 80 |
| 170318A | 12:11:56 | $20^{h}22^{m}39^{s}$ | 28°24′ | *Swift* | Long | H1L1$^{\dagger}$ | ⋯ | ⋯ | ⋯ | 47 | 119 |
| 170318B | 15:27:52 | $18^{h}57^{m}10^{s}$ | 6°19′ | *Swift* | Short | H1L1 | 152 | 254 | 281 | 48 | 112 |
| 170323058 | 01:23:23 | $9^{h}40^{m}45^{s}$ | −38°60′ | *Fermi* | Long | H1L1 | ⋯ | ⋯ | ⋯ | 28 | 75 |
| 170325331 | 07:56:58 | $8^{h}29^{m}55^{s}$ | 20°32′ | *Fermi* | Short | H1L1 | 73 | 88 | 125 | 33 | 77 |
| 170330A | 22:29:51 | $18^{h}53^{m}17^{s}$ | −13°27′ | *Swift* | Long | H1L1$^{\dagger}$ | ⋯ | ⋯ | ⋯ | 41 | 110 |
| 170331A | 01:40:46 | $21^{h}35^{m}06^{s}$ | −24°24′ | *Swift* | Long | H1L1 | ⋯ | ⋯ | ⋯ | 49 | 119 |
| 170402285 | 06:50:54 | $22^{h}01^{m}26^{s}$ | −10°38′ | *Fermi* | Long | H1L1 | ⋯ | ⋯ | ⋯ | 9 | 110 |
| 170402961 | 23:03:25 | $20^{h}31^{m}40^{s}$ | −45°56′ | *Fermi* | Long | H1L1 | ⋯ | ⋯ | ⋯ | 48 | 113 |
| 170403583 | 13:59:18 | $17^{h}\,48^{m}19^{s}$ | 14°31′ | *Fermi* | Short | H1L1 | 166 | 240 | 261 | ⋯ | ⋯ |
| 170403707 | 16:57:33 | $16^{h}24^{m}09^{s}$ | 41°49′ | *Fermi* | Long | H1L1 | ⋯ | ⋯ | ⋯ | 24 | 54 |
| 170409112 | 02:42:00 | $23^{h}10^{m}19^{s}$ | −7°04′ | *Fermi* | Long | H1L1$^{\dagger}$ | ⋯ | ⋯ | ⋯ | 20 | 106 |
| 170414551 | 13:13:16 | $2^{h}54^{m}00^{s}$ | 75°53′ | *Fermi* | Long | H1L1 | ⋯ | ⋯ | ⋯ | 33 | 80 |
| 170416583 | 14:00:05 | $18^{h}56^{m}52^{s}$ | −57°01′ | *Fermi* | Long | H1L1$^{\dagger}$ | ⋯ | ⋯ | ⋯ | 9 | 24 |
| 170419983 | 23:36:14 | $17^{h}39^{m}28^{s}$ | −11°14′ | *Fermi* | Long | H1L1 | ⋯ | ⋯ | ⋯ | 49 | 119 |
| 170419A | 13:26:40 | $5^{h}19^{m}25^{s}$ | −21°26′ | *Swift* | Long | H1L1 | ⋯ | ⋯ | ⋯ | 48 | 114 |
| 170422343 | 08:13:54 | $12^{h}34^{m}31^{s}$ | 16°49′ | *Fermi* | Long | H1L1 | ⋯ | ⋯ | ⋯ | 47 | 114 |
| 170423719 | 17:15:08 | $22^{h}57^{m}21^{s}$ | −4°16′ | *Fermi* | Long | H1L1 | ⋯ | ⋯ | ⋯ | 36 | 98 |
| 170423872 | 20:55:23 | $13^{h}58^{m}24^{s}$ | 26°22′ | *Fermi* | Long | H1L1 | ⋯ | ⋯ | ⋯ | 17 | 45 |
| 170424 | 10:12:06 | $10^{h}00^{m}40^{s}$ | −13°41′ | IPN | Long | H1L1 | ⋯ | ⋯ | ⋯ | 32 | 75 |
| 170424425 | 10:12:30 | $22^{h}54^{m}07^{s}$ | −45°12′ | *Fermi* | Long | H1L1 | ⋯ | ⋯ | ⋯ | 32 | 74 |
| 170428136 | 03:16:17 | $0^{h}19^{m}02^{s}$ | 56°14′ | *Fermi* | Long | H1L1 | ⋯ | ⋯ | ⋯ | 23 | 75 |
| 170428A[e] | 09:13:42 | $22^{h}00^{m}12^{s}$ | 26°55′ | *Swift* | Short | H1L1 | 105 | 167 | 178 | 32 | 86 |
| 170430204 | 04:54:20 | $01^{h}\,35^{m}26^{s}$ | 30°07′ | *Fermi* | Short | H1 | 32 | 54 | 81 | ⋯ | ⋯ |
| 170501467 | 11:11:53 | $6^{h}28^{m}02^{s}$ | 13°43′ | *Fermi* | Long | H1L1 | ⋯ | ⋯ | ⋯ | 34 | 84 |
| 170506169 | 04:02:48 | $7^{h}29^{m}02^{s}$ | 51°52′ | *Fermi* | Ambiguous | H1L1 | 103 | 174 | 149 | 36 | 84 |
| 170604603 | 14:28:05 | $22^{h}\,41^{m}36^{s}$ | 40°42′ | *Fermi* | Short | L1 | 131 | 204 | 237 | ⋯ | ⋯ |
| 170610689 | 16:31:47 | $4^{h}35^{m}38^{s}$ | 46°29′ | *Fermi* | Long | H1L1 | ⋯ | ⋯ | ⋯ | 53 | 162 |
| 170611937 | 22:29:35 | $11^{h}34^{m}19^{s}$ | −7°22′ | *Fermi* | Long | H1L1 | ⋯ | ⋯ | ⋯ | 32 | 75 |
| 170614255 | 06:06:41 | $4^{h}42^{m}12^{s}$ | 37°56′ | *Fermi* | Long | H1L1$^{\dagger}$ | ⋯ | ⋯ | ⋯ | 22 | 55 |

**Table 3**
(Continued)

| | | | | | | | $D_{90}$ (Mpc) | | | | |
|---|---|---|---|---|---|---|---|---|---|---|---|
| GRB Name | UTC Time | R.A. | Decl. | Satellite(s) | Type | Network | BNS | Generic NS–BH | Aligned NS–BH | ADI-A | CSG-150 Hz |
| 170614505 | 12:06:39 | $20^h\,43^m58^s$ | −37°54′ | *Fermi* | Ambiguous | H1 | 9 | 22 | 0 | … | … |
| 170616165 | 03:58:07 | $3^h18^m02^s$ | 19°40′ | *Fermi* | Long | H1L1† | … | … | … | 34 | 95 |
| 170618475 | 11:24:41 | $0^h59^m19^s$ | 26°44′ | *Fermi* | Long | H1L1 | … | … | … | 48 | 130 |
| 170625692 | 16:35:47 | $7^h06^m48^s$ | −69°21′ | *Fermi* | Long | H1L1 | … | … | … | 33 | 84 |
| 170626A | 09:37:23 | $11^h01^m37^s$ | 56°29′ | *Swift* | Long | H1L1 | … | … | … | 33 | 82 |
| 170629A | 12:53:33 | $8^h39^m50^s$ | −46°35′ | *Swift* | Long | H1L1 | … | … | … | 48 | 117 |
| 170705200 | 04:48:30 | $23^h58^m02^s$ | −21°56′ | *Fermi* | Long | H1L1 | … | … | … | 29 | 74 |
| 170705244 | 05:50:45 | $15^h50^m26^s$ | −7°26′ | *Fermi* | Long | H1L1 | … | … | … | 32 | 86 |
| 170705A[f] | 02:45:47 | $12^h46^m50^s$ | 18°18′ | *Swift* | Long | H1L1† | … | … | … | 47 | 156 |
| 170708046 | 01:06:11 | $22^h\,13^m00^s$ | 25°37′ | *Fermi* | Short | L1 | 57 | 105 | 103 | … | … |
| 170709334 | 08:00:24 | $20^h\,40^m10^s$ | 02°12′ | *Fermi* | Ambiguous | L1 | 139 | 228 | 255 | … | … |
| 170714A[g] | 12:25:32 | $2^h17^m17^s$ | 1°58′ | *Swift* | Long | H1L1† | … | … | … | 48 | 123 |
| 170715878 | 21:04:13 | $19^h08^m52^s$ | −16°37′ | *Fermi* | Long | H1L1 | … | … | … | 47 | 114 |
| 170723076 | 01:49:10 | $9^h03^m45^s$ | −19°26′ | *Fermi* | Long | H1L1 | … | … | … | 26 | 75 |
| 170723677 | 16:15:27 | $1^h28^m16^s$ | 62°41′ | *Fermi* | Long | H1L1 | … | … | … | 37 | 111 |
| 170723882 | 21:10:18 | $14^h10^m19^s$ | 39°50′ | *Fermi* | Ambiguous | H1L1 | 95 | 83 | 179 | 40 | 110 |
| 170724A | 00:48:44 | $10^h00^m14^s$ | −1°02′ | *Swift* | Long | H1L1† | … | … | … | 21 | 84 |
| 170726249 | 05:58:15 | $11^h05^m40^s$ | −34°00′ | *Fermi* | Ambiguous | H1L1 | 124 | 152 | 207 | 38 | 112 |
| 170728A | 06:53:28 | $3^h55^m36^s$ | 12°10′ | *Swift* | Short | H1L1 | 89 | 129 | 163 | 26 | 81 |
| 170731751 | 18:01:39 | $16^h20^m48^s$ | 64°18′ | *Fermi* | Long | H1L1† | … | … | … | 17 | 44 |
| 170802638 | 15:18:24 | $3^h29^m12^s$ | −39°13′ | *Fermi* | Ambiguous | H1L1V1 | 45 | 62 | 72 | 3 | 24 |
| 170803172 | 04:07:15 | $5^h06^m00^s$ | 23°60′ | *Fermi* | Ambiguous | H1L1 (H1L1V1)[h] | 56 | 83 | 105 | 16 | 53 |
| 170803B | 22:00:32 | $00^h\,56^m53^s$ | 06°34′ | IPN | Short | L1[i] | 140 | 215 | 234 | … | … |
| 170804A | 12:01:37 | $0^h25^m37^s$ | −64°47′ | *Swift* | Long | H1V1† | … | … | … | 15 | 45 |
| 170805901 | 21:37:49 | $16^h15^m52^s$ | 36°23′ | *Fermi* | Long | H1V1 | … | … | … | 11 | 25 |
| 170805A | 14:38:10 | $20^h50^m26^s$ | 22°28′ | IPN | Short | H1L1V1 | 69 | 100 | 114 | 22 | 61 |
| 170805B | 14:18:49 | $8^h40^m32^s$ | 70°06′ | IPN | Short | H1L1V1 | 132 | 163 | 218 | 33 | 114 |
| 170807A | 21:56:09 | $9^h33^m44^s$ | −17°21′ | *Swift* | Long | H1L1 | … | … | … | 27 | 76 |
| 170808065 | 01:34:09 | $0^h13^m12^s$ | 62°18′ | *Fermi* | Ambiguous | L1V1 | 58 | 83 | 87 | 11 | 18 |
| 170808936 | 22:27:43 | $9^h42^m38^s$ | 2°11′ | *Fermi* | Long | L1V1 | … | … | … | 22 | 41 |
| 170809 | 23:46:26 | $16^h52^m37^s$ | −12°18′ | IPN | Long | H1L1V1 | … | … | … | 27 | 87 |
| 170816258 | 06:11:11 | $0^h42^m48^s$ | −15°37′ | *Fermi* | Long | H1L1† | … | … | … | 17 | 55 |
| 170816599 | 14:23:03 | $23^h25^m36^s$ | 19°06′ | *Fermi* | Short | H1L1V1 (H1V1)[j] | 46 | 56 | 73 | 15 | 34 |
| 170817908 | 21:47:34 | $5^h32^m07^s$ | 50°04′ | *Fermi* | Ambiguous | H1V1 | 35 | 51 | 63 | 16 | 30 |
| 170817A | 12:41:06 | $13^h\,09^m36^s$ | −23°24′ | *Fermi* | Ambiguous | H1L1V1 | N/A | N/A | N/A | N/A | N/A |
| 170818137 | 03:17:20 | $19^h\,48^m53^s$ | 06°21′ | *Fermi* | Ambiguous | H1L1 | 103 | 146 | 169 | … | … |
| 170821265 | 06:22:00 | $16^h51^m26^s$ | 19°07′ | *Fermi* | Long | H1L1† | … | … | … | 33 | 76 |
| 170822A | 09:11:51 | $6^h17^m29^s$ | 54°60′ | *Swift* | Long | H1L1V1† | … | … | … | 32 | 97 |
| 170823A | 22:16:48 | $12^h34^m51^s$ | 35°33′ | *Swift* | Long | H1L1† | … | … | … | 58 | 166 |
| 170825307 | 07:22:01 | $18^h17^m36^s$ | −26°12′ | *Fermi* | Long | L1V1 | … | … | … | 15 | 31 |
| 170825500 | 12:00:06 | $0^h14^m33^s$ | 20°07′ | *Fermi* | Long | H1L1 | … | … | … | 47 | 116 |
| 170825784 | 18:49:11 | $7^h45^m16^s$ | −48°43′ | *Fermi* | Long | H1L1V1† | … | … | … | 6 | 22 |

**Notes.** The "Satellite(s)" column lists the instrument whose sky localization was used for the purposes of analysis. The "Network" column lists the GW detector network used in the analysis of each GRB—H1 = LIGO Hanford; L1 = LIGO Livingston; V1 = Virgo. A dagger denotes cases in which the on-source window of the generic transient search is extended to cover the GRB duration ($T_{90} > 60$ s). In cases where each analysis used a different network, parentheses indicate the network used for `PyGRB` analysis, and detail is provided in the table footnotes. Columns (8)–(12) display the 90% confidence exclusion distances to the GRB ($D_{90}$) for several emission scenarios: BNS, generic and aligned-spin NS–BH, ADI-A, and CSG GW burst at 150 Hz with total radiated energy $E_{\rm GW} = 10^{-2}\,M_\odot c^2$.

[a] GRB 170113A has a redshift of $z = 1.968$ (Xu et al. 2017).
[b] GRB 170125102 occurred when the Livingston detector was not in its nominal observing state; however, the data were deemed suitable for the purposes of the unmodeled analysis.
[c] GRB 170202A has a redshift of $z = 3.645$ (de Ugarte Postigo et al. 2017a).
[d] GRB 170305256 occurred near the null of the Hanford detector, and inclusion of its data degraded the `PyGRB` search sensitivity compared to a Livingston-only analysis.
[e] GRB 170428A has a redshift of $z = 0.454$ (Izzo et al. 2017).
[f] GRB 170705A has a redshift of $z = 2.01$ (de Ugarte Postigo et al. 2017b).
[g] GRB 170714A has a redshift of $z = 0.793$ (de Ugarte Postigo et al. 2017c).
[h] GRB 170803172: Virgo data did not meet the data quality requirements of `X-Pipeline`.
[i] GRB 170803B occurred near the null of the Virgo detector (see note *b*). In addition, Livingston data did not meet the data quality requirements of `X-Pipeline`, so this GRB was not subject to the unmodeled analysis.
[j] GRB 170816599 occurred near the null of the Livingston detector (see note b).

increases by only a factor of about two. This discrepancy highlights the fact that faint, wide-angle emission will remain detectable for only nearby mergers, meaning that additional joint GW BNS detections facilitated by improved GW detector sensitivity will require the system to have small inclinations in order to produce a detectable GRB.

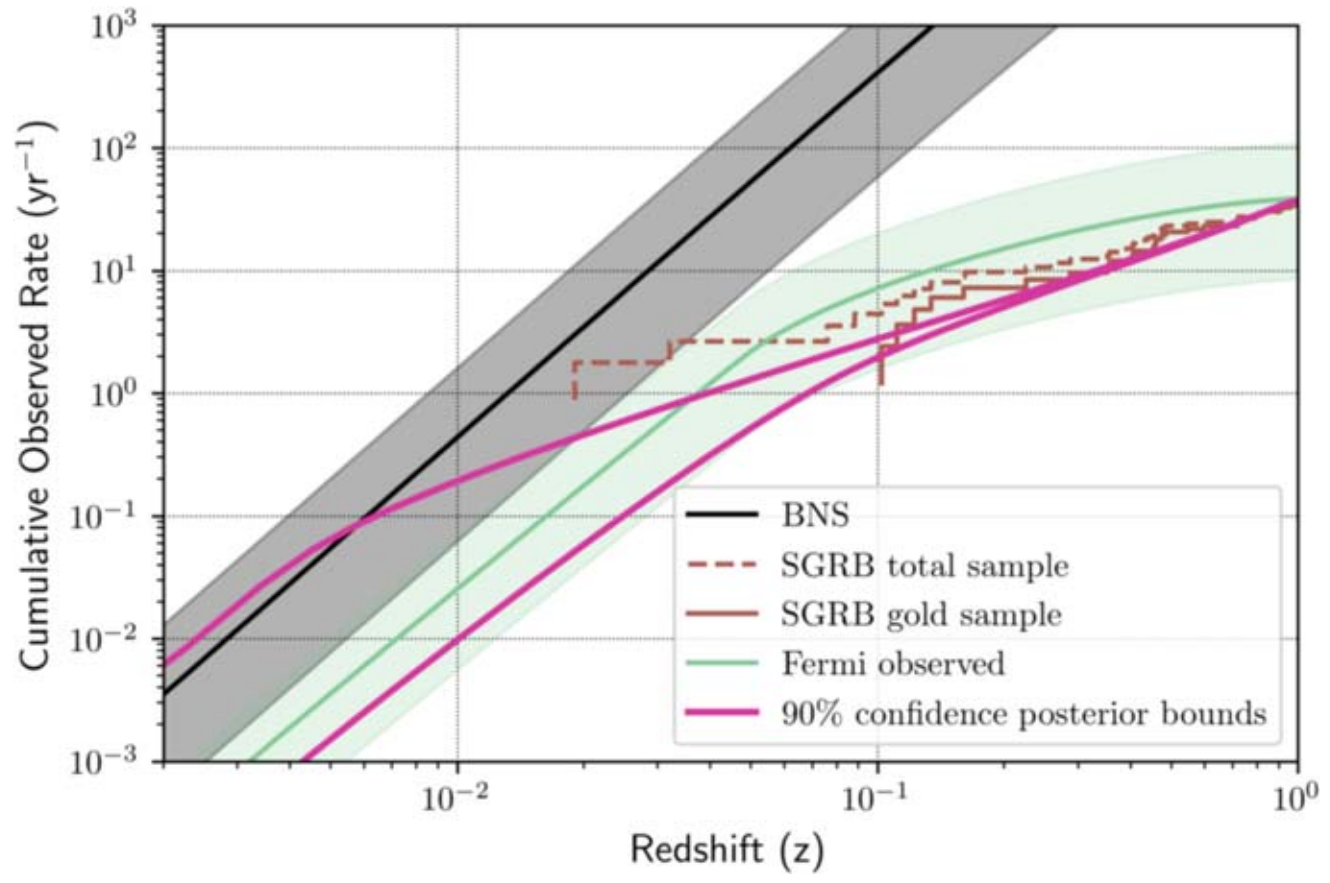

**Figure 5.** Predicted event rates per year as a function of redshift. The magenta lines show the 90% bounds on the rate associated with the fit of our model of the short-GRB luminosity function (Equation (1)) to the O2 run results. In black we show the BNS merger rate $1210^{+3230}_{-1040}$ Gpc$^{-3}$ yr$^{-1}$ (Abbott et al. 2019a), and in green we show the *Fermi*-GBM short-GRB detection rate and its 90% credible interval (Howell et al. 2019). As a reference, the measured short-GRB redshift distribution without GRB 170817A is shown in brown (Abbott et al. 2017e, and references therein). Our analysis results, shown in magenta, are compatible with the BNS merger rate and the *Fermi*-GBM observed short-GRB rate. This is consistent with the hypothesis that BNS mergers are generally short-GRB progenitors.

## 7. Conclusions

We have performed targeted analyses for GWs in association with GRBs during O2, searching for NS binary merger signals from short GRBs with a modeled analysis and GW burst signals from all GRBs with an unmodeled analysis. GW170817 is confirmed by both methods as a strong detection associated with GRB 170817A, entirely consistent with previously published results. No further GW signals were found as a result of these analyses, and there is no strong evidence found in our results for subthreshold signals. We set lower bounds on the distances to progenitors for a number of emission models, which include the largest $D_{90}$ values published so far for some individual GRBs (Abadie et al. 2012a; Abbott et al. 2017g).

Based on the results of the modeled search, we performed a population model analysis in Section 6 and place bounds on a twice-broken power-law short-GRB luminosity function that is consistent with both the measured BNS merger rate and the *Fermi*-GBM observed short-GRB rate, and therefore with the hypothesis that BNS mergers are generally short-GRB progenitors. Further multimessenger observations should provide tighter constraints on GRB emission models and event rates and investigate whether NS–BH mergers also power short GRBs. We expect to observe 0.07–1.80 joint GRB-GW events per year in conjunction with *Fermi*-GBM during the 2019–2020 LIGO-Virgo observing run and 0.15–3.90 per year when GW detectors are operating at their design sensitivities.

The authors gratefully acknowledge the support of the United States National Science Foundation (NSF) for the construction and operation of the LIGO Laboratory and Advanced LIGO, as well as the Science and Technology Facilities Council (STFC) of the United Kingdom, the Max-Planck-Society (MPS), and the State of Niedersachsen/Germany for support of the construction of Advanced LIGO and construction and operation of the GEO600 detector. Additional support for Advanced LIGO was provided by the Australian Research Council. The authors gratefully acknowledge the Italian Istituto Nazionale di Fisica Nucleare (INFN), the French Centre National de la Recherche Scientifique (CNRS), and the Foundation for Fundamental Research on Matter supported by the Netherlands Organisation for Scientific Research, for the construction and operation of the Virgo detector and the creation and support of the EGO consortium. The authors also gratefully acknowledge research support from these agencies, as well as by the Council of Scientific and Industrial Research of India; the Department of Science and Technology, India; the Science & Engineering Research Board (SERB), India; the Ministry of Human Resource Development, India; the Spanish Agencia Estatal de Investigación; the Vicepresidència i Conselleria d'Innovació Recerca i Turisme and the Conselleria d'Educació i Universitat del Govern de les Illes Balears; the Conselleria d'Educació Investigació Cultura i Esport de la Generalitat Valenciana; the National Science Centre of Poland; the Swiss National Science Foundation (SNSF); the Russian Foundation for Basic Research; the Russian Science Foundation; the European Commission; the European Regional Development Funds (ERDF); the Royal Society; the Scottish Funding Council; the Scottish Universities Physics Alliance; the Hungarian Scientific Research Fund (OTKA); the Lyon Institute of Origins (LIO); the Paris Île-de-France Region; the National Research, Development and Innovation Office Hungary (NKFIH); the National Research Foundation of Korea; Industry Canada and the Province of Ontario through the Ministry of Economic Development and Innovation; the Natural Science and Engineering Research Council Canada; the Canadian Institute for Advanced Research; the Brazilian Ministry of Science, Technology, Innovations, and Communications; the International Center for Theoretical Physics South American Institute for Fundamental Research (ICTP-SAIFR); the Research Grants Council of Hong Kong; the National Natural Science Foundation of China (NSFC); the Leverhulme Trust; the Research Corporation; the Ministry of Science and Technology (MOST), Taiwan; and the Kavli Foundation. The authors gratefully acknowledge the support of the NSF, STFC, INFN, and CNRS for provision of computational resources. D.S.S., D.D.F., R.L.A., and A.V.K. acknowledge support from RSF grant 17-12-01378.

*Facilities:* LIGO, EGO:Virgo, *Fermi* (GBM), *Swift* (BAT), *INTEGRAL*, *WIND* (KONUS), *Odyssey*.

*Software:* `Matplotlib` (Hunter 2007; Caswell et al. 2018), `LALInference` (Veitch et al. 2015), `PyCBC` (Nitz et al. 2018), `X-Pipeline` (Sutton et al. 2010; Was et al. 2012).

## ORCID iDs

K. Ackley https://orcid.org/0000-0002-8648-0767
L. P. Singer https://orcid.org/0000-0001-9898-5597
M. Spera https://orcid.org/0000-0003-0930-6930
Shubhanshu Tiwari https://orcid.org/0000-0003-1611-6625
K. Ueno https://orcid.org/0000-0003-0424-3045
S. Vitale https://orcid.org/0000-0003-2700-0767
J. T. Whelan https://orcid.org/0000-0001-5710-6576
M. Zevin https://orcid.org/0000-0002-0147-0835
D. D. Frederiks https://orcid.org/0000-0002-1153-6340
D. S. Svinkin https://orcid.org/0000-0002-2208-2196
Badri Krishnan https://orcid.org/0000-0003-3015-234X